\documentclass[aps,prd,twocolumn,groupedaddress,showpacs,preprintnumbers]{revtex4}

\usepackage{graphicx}% Include figure files
\usepackage{color}%for preprints

\begin{document}


\preprint{ICRR-Report-600-2011-17}

\title{Resonant annihilation of long-lived massive colored particles through hadronic collisions}


\author{Motohiko Kusakabe$^{1}$\footnote{kusakabe@icrr.u-tokyo.ac.jp} and Tomohiro Takesako$^{1,2}$}

%\email{kusakabe@icrr.u-tokyo.ac.jp}
%\thanks{}
\affiliation{
$^1$Institute for Cosmic Ray Research, University of Tokyo, Kashiwa,
Chiba 277-8582, Japan\\
$^2$Department of Physics, Graduate School of Science, University of Tokyo, Hongo, Bunkyo-ku, Tokyo 113-0033, Japan}


\date{\today}

\begin{abstract}
Hypothetical long-lived massive colored particles (MCPs or $Y$s) would be confined in colorless exotic strongly interacting massive particles (SIMPs) at color confinement temperature of $T_{\rm C}\sim 200$~MeV.  Two long-lived MCPs form a bound state $(Y\bar{Y})$ at collisions of two SIMPs.  We study sensitivities of MCP annihilation to decay properties of resonances $(Y\bar{Y})$, and binding energies or energy levels of exotic SIMPs.  The $(Y\bar{Y})$ formation is assumed to dominantly proceed through resonances of $(Y\bar{Y})$ in this paper.  We make a toy model of the effective cross section for $Y\bar{Y}$ annihilation.  Abundances of SIMPs are then calculated for different sets of parameters specifying properties of $(Y\bar{Y})$ resonances, binding energies of SIMPs, the initial abundance and the mass of MCP.  Calculated relic abundances for respective SIMP species are $2\times 10^{-8}$--$3\times 10^{-4}$ times that of baryon.  They can be much higher but cannot be much smaller than the previous estimate.
The abundances can be consistent depending on parameters with the possible scenario that SIMPs bind to nuclei and subsequent exotic nuclear reactions reduce the primordial abundance of $^7$Li or enhance those of $^9$Be and/or B in the early Universe.  A unique information on the quark-hadron phase transition in the early Universe may become available in future by elaborated studies on the annihilation process with light element abundances as observables.
\end{abstract}

%98.80.Cq 	Particle-theory and field-theory models of the early Universe (including cosmic pancakes, cosmic strings, chaotic phenomena, inflationary universe, etc.)
%14.80.Pq 	R-hadrons
%13.75.-n 	Hadron-induced low- and intermediate-energy reactions and scattering (energy ≤ 10 GeV)
%98.80.Es 	Observational cosmology (including Hubble constant, distance scale, cosmological constant, early Universe, etc)

%\keywords{}

\maketitle

\section{Introduction}\label{sec1}

In some particle models beyond the standard model (SM), long-lived massive colored particles (MCPs) can exist.  The models include a long-lived gluino in split supersymmetry~\cite{ArkaniHamed:2004fb,ArkaniHamed:2004yi}, the next-to-lightest supersymmetric particle (NLSP) gluino~\cite{Raby:1997bpa,Shirai:2010rr,Covi:2010au}, and NLSP squark~\cite{Sarid:1999zx} both in a weak scale supersymmetry.  In addition, extended theories with other kinds of colored particles have been considered~\cite{Hisano:2010bx,Nakayama:2010vs}.  Such models will be tested at weak scale in the Large Hadron Collider (LHC).

If long-lived MCPs exist at the temperature of color deconfinement, i.e., $T_{\rm C}\sim 200$~MeV, they would be confined inside exotic heavy hadrons, so-called strongly interacting massive particles (SIMPs, or $X$ particles)~\cite{Wolfram:1978gp,Dover:1979sn,Starkman:1990nj}.

Dover, Gaisser \& Steigman \cite{Dover:1979sn} have originally estimated the relic abundance of SIMPs, and studied the possibility of exotic stable hadrons with limits from terrestrial superheavy element searches.  They adopted the thermal rate of MCP annihilation in collisions of SIMPs which is assumed to be the same as (or less than) that for nucleons, and $\sigma v\sim 30$~mb.  The final abundance thus estimated was about $10^{-11}$ times the baryon abundance.

Nardi \& Roulet \cite{Nardi1990} have studied the possibility of new stable exotic quarks predicted by the E$_6$ model.  They concluded that the possibility is ruled out unless there is some mechanism for decay of such quarks.  Also, they pointed out that the annihilation cross section used in Ref. \cite{Dover:1979sn} is significantly overestimated.  The first reason is that the cross section includes extra final states irrelevant to the MCP annihilation.  The second is the de Broglie wave length of a two-MCP system is the maximum radius scale for the total annihilation cross section.  It is never larger than the interaction radius of $\sim$1~fm until the temperature decreases down to $\sim$10 MeV~$(m/{\rm 1~TeV})^{-1}$ with $m$ the mass ($m\gg 1$~GeV) of the MCPs.  However, the previous work~\cite{Dover:1979sn} used a cross section larger than $1~{\rm fm}^{-2}$ above the temperature.

Thermal relic abundances after the freeze-out of annihilations in the early Universe have been studied with various assumptions on annihilation cross sections~\cite{Baer:1998pg}.  Theoretical estimates, however, predicted various values ranging over more than several order of magnitude depending on adopted cross sections~\cite{Baer:1998pg}.

Kang, Luty and Nasri~\cite{Kang:2006yd} studied aspects of annihilation of SIMPs at $T\sim T_{\rm C}$.  They showed that the annihilation proceeds through the following steps: the formation of bound states of two MCPs with large orbital angular momentums, their transitions to more bound lower-energy levels, and the annihilation of MCPs.  They suggested that the bound state formation can realize from multiple partial waves of initial relative angular momenta in the two-SIMP system.  They then suppose a geometrical cross section for annihilation at temperatures $T\sim T_{\rm C}$.

The final abundance of SIMP has been estimated~\cite{Kang:2006yd} to be
\begin{equation}
\frac{N_X}{s} \sim \sqrt[]{\mathstrut \frac{15}{\pi}}
 \frac{g_\ast^{1/2}}{g_{\ast s}} \frac{m^{1/2}}{\sigma T_B^{3/2} m_{\rm Pl}}~~,
\label{eq1}
\end{equation}
where
$N_X$ is the number density of SIMP,
$s=2\pi^2 g_{\ast s} T^3/45$ is the entropy density with $g_{\ast s}\sim 10$ the total number of
effective massless degrees of freedom in terms of entropy~\cite{kolb1990} below the phase transition of quantum chromodynamics (QCD),
$g_\ast$ is the total number of effective massless degrees of
freedom in number~\cite{kolb1990},
$\sigma$ is the annihilation cross section of the SIMP,
$T_B$ is the temperature of the Universe at which the colorless SIMPs are
formed,
and $m_{\rm Pl}$ is the Planck mass.
The ratio of number abundance of SIMP  to that of baryons is then ${N_X}/{n_b}\sim 0.5\times 10^{-8}$, where $n_b$ is the number density of baryons.

Diaz-Cruz et al.~\cite{DiazCruz:2007fc} have adopted models in which the lighter stop $\tilde{t_1}$ is the NLSP and the gravitino is the LSP.  They studied cosmological behaviors of the stop, and find that stops are finally hadronized into mesinos or sbaryons.  They used data on energy levels of normal hadron and the estimation of annihilation rate by Kang et al. \cite{Kang:2006yd}, and showed an abundance evolution of exotic hadrons.

Jacoby \& Nussinov~\cite{Jacoby:2007nw,Nussinov:2009hc} have refined the estimate of relic abundance in Ref.~\cite{Kang:2006yd}.  They assumed that MCPs are quarklike, and considered a reaction between SIMPs and normal hadrons, i.e., mesons and nucleons, which transform constituents or energy levels of exotic hadrons $X$.  They suggested that the dissociation of bound states of two MCPs must be considered.  They roughly took all those effects into account, and concluded that the relic abundance after the freeze-out of the annihilation at $T\lesssim T_{\rm C}$ is as small as suggested in Ref.~\cite{Kang:2006yd}.

Recently the ''Quirk Model'' has been suggested by M. Luty~\cite{Jacoby:2007nw}.  In the model, colored particles transform under an additional non-Abelian gauge group.  The phenomenology relating to a new exotic quanta of the corresponding gauge field originally named theton had been studied about three decades ago~\cite{Okun:1980kw,Okun1980b}.  They have discussed results for the situation in which the mass of the lightest quark was heavier than the QCD parameter $\Lambda_{\rm had}\sim 1$~GeV~\cite{Gupta:1981ve}. Taking account of the recent suggestions~\cite{Kang:2006yd,Jacoby:2007nw}, an estimate of relic abundances has derived, and collider phenomenology of such particles or the possibility of detection at LHC or Tevatron has been studied~\cite{Jacoby:2007nw,Kang:2008ea}.  Strassler and Zurek~\cite{Strassler:2006im} have also studied the phenomenology of a new confining gauge group added to SM as a kind of hidden-valley models.

In this paper, we construct a toy model which roughly includes effects of all possible processes operating in the annihilation at $T\lesssim T_{\rm C}$.  Abundances of SIMPs during the annihilation epoch are calculated for four cases of different properties of resonant or barely-bound states of two-MCP systems and binding energies of SIMPs.  We show results for each case with different initial abundances and masses of MCPs.  We then suggest that final abundances can be much higher but cannot be much smaller than the naive estimations of Refs. \cite{Kang:2006yd,Jacoby:2007nw}.

In Sec.~\ref{sec2}, we describe input physics and assumptions adopted in this paper.  We show a hypothetical quantum mechanical picture of annihilation of SIMPs (Sec.~\ref{sec2}A), a toy model of the effective cross section for the annihilation (Sec.~\ref{sec2}B), a comparison of the formation rate of bound states involving two MCPs with the cosmic expansion rate (Sec.~\ref{sec2}C), the transformation of mesonlike SIMPs to nucleonlike SIMPs (Sec.~\ref{sec2}D), and the annihilation cross section at low temperature after the transformation (Sec.~\ref{sec2}E).  In Sec.~\ref{sec3}, we describe models of different decay properties of resonances of two MCPs, and different binding energies of SIMPs.  Effective annihilation cross sections for the models are derived.  In Sec.~\ref{sec4}, we show result of calculations of SIMP abundance and effects of decay properties of resonances and binding energies.  In Sec.~\ref{sec5}, we briefly mention the case in which heavy hadrons including two MCPs exist stably.  In Sec.~\ref{sec6}, this work is summarized, and a possible impact of results about relic abundance of SIMPs on big bang nucleosynthesis is discussed.

In this paper, the following units are adopted: the Planck's constant, the Boltzmann's constant, and the light speed are unity, i.e., $\hbar=1$, $k_{\rm B}=1$ and $c=1$.

\section{Input physics}\label{sec2}
\subsection{A quantum mechanical picture of resonant $(Y\bar{Y})$ production}
Massive colored particles (MCPs), i.e., $Y$, would be confined in colorless states, i.e., $X$, at $T_{\rm C}\sim 200$~MeV~\cite{Kang:2006yd}.  Such states can react with each other and transform into a bound state via reactions of the type
\begin{equation}
X_1+X_2 \rightarrow (Y\bar{Y}) + \{\gamma~{\rm or}~q\bar{q}~{\rm or}~qqq~...\},
\label{eq2}
\end{equation}
where
$X_1$ and $X_2$ are states including one $Y$ or $\bar{Y}$ particle,
$(Y\bar{Y})$ is a bound state including one $Y$ and one $\bar{Y}$ particle.

The Schr$\ddot{\rm o}$dinger equation for a two-body wave function, $\chi_l(r)=r \psi_l(r)$ is given by
\begin{equation}
\left[-\frac{1}{2\mu} \frac{d^2}{dr^2} +\frac{l(l+1)}{2\mu r^2} +V(r)-E \right]\chi_l(r)=0,
\label{eq3}
\end{equation}
where
$\mu$ is the reduced mass, 
$r$ is the radius of the system,
$V(r)$ is the central potential at $r$, 
$E$ is the kinetic energy,
and $\psi(r)$ is the wave function at $r$. 
The centrifugal potential for an initial state ($X_1$+$X_2$) is small because of its large reduced mass.  The penetration factor can, therefore, be large even for high angular momentum $l$.

Kang et al.~\cite{Kang:2006yd} suggested that cross sections for $X_1+X_2$ reaction [Eq. (\ref{eq2})] would be as large as that for reactions of normal hadrons with their typical values of $\sim [\pi (1~{\rm GeV}^{-1})^{2}]$.  We note that the cross sections can be smaller than their assumption according to a well-known concept from nuclear reactions as follows:

Low-lying resonances or excited states of $(Q\bar{Q})^\ast$ with heavy quarks $Q$, such as $c\bar{c}$ or $b\bar{b}$, marginally bound with respect to the $Q\bar{q}+\bar{Q}q$ channel have large width generally.   The reaction for exotic MCPs is then also considered to proceed through such states, i.e., $(Y\bar{Y})^\ast$.  

The reaction [Eq. (\ref{eq2})] for normal nucleons, i.e., $N$, is $N_1+N_2\rightarrow (NN)+\pi$.  This is an important reaction of pion production and its cross section is measured \cite{Richard1970}.  The cross section is theoretically explained assuming that the reaction has a contribution from $\Delta(1232)$ resonance ($N^\ast$) composed of a pion and a nucleon and the existence of state $NN^\ast$ \cite{Mandelstam1958,Schiff1968}.  Such a situation is assumed in the reaction [Eq. (2)] although the reaction mechanism is expected to be much more complicated by many partial waves playing roles and many resonant or bound energy levels of $(Y\bar{Y})$ as described below.

A nonresonant component of the reaction might significantly contribute to the total cross section.  In this study, however, we assume that the reaction operates dominantly through resonances~\footnote{For cases of charmonia and bottomonia, dissociations of the bound states at collisions with pions, which are inverse reactions of Eq. (\ref{eq2}), have been studied.  The consideration of such reactions and a temperature dependence of hadronic energy levels~\cite{Wong:2001uu} is beyond the scope of this study.  They should be included in the future.}.  The cross section would then be described using the Breit-Wigner formula~\cite{clayton-book}, i.e.,
\begin{equation}
\sigma=\frac{\pi}{2\mu E} \sum_i \omega_i \frac{\Gamma_{i,X} \Gamma_{i,L}}{(E-E_i)^2+(\Gamma_i/2)^2},
\label{eq4}
\end{equation}
where
$E$ is the kinetic energy in the entrance channel.
The sum is taken over resonances $i$.
$\omega_i=g_i/(g_{X_1} g_{X_2})$ is the statistical weight factor with $g_a=2J_a+1$ the spin ($J_a$) degeneracy factors for $a=i$, i.e., the resonant or excited state, heavy hadrons $X_1$ and $X_2$, respectively.

The total decay width of the state $i$ is given by
\begin{equation}
\Gamma_i=\sum_j \Gamma_{i,j}= \Gamma_{i,L}+\Gamma_{i,X},
\label{eq5}
\end{equation}
where
$\Gamma_{i,j}$ is the partial width for decay of $i$ to $j$.  It is given by $\Gamma_{i,j}=\sum_l \Gamma_{i,j,l}$ with $\Gamma_{i,j,l}$ the partial width for decay of $i$ to $j$ of relative angular momentum $l$.
$\Gamma_{i,L}= \Gamma_{i,\gamma}+ \Gamma_{i,\pi\pi}+\Gamma_{i,\pi}+...$ is the decay width of $i$ into all light particles ($L$) composed of SM particles.  $L$ does not includes $X_1$ and $X_2$.

In nuclear reactions, the partial width for emissions of nucleon or nuclei is expressed by
\begin{equation}
\Gamma_{i,j,l}=\frac{3v_{i,j}}{R_{i,j}} P_{i,j,l} \theta_{i,j,l}^2,
\label{eq6}
\end{equation}
where
$v_{i,j}$ is the velocity of the exit channel $j$ in reaction through the state $i$,
$R_{i,j}$ is hadronic interaction radius,
$P_{i,j,l}$ is the penetration factor for particles ($j$) of relative angular momentum $l$.
The quantity $\theta_{i,j,l}^2$ is the dimensionless reduced width which means the probability that the state $i$ is described by the state of $j$ of angular momentum $l$.  The reduced width is empirically known to take values of $0.01<\theta_{i,j,l}<1$ for nuclear states $i$~\cite{clayton-book}.

There are lots of decay modes of hadrons in which light particles are emitted.  As we describe hereinbelow, however, existing experimental data on heavy quarkonia \cite{PDG2010} show that the radiative decay width is predominant over others.  We, therefore, assume that the width for decay into all light particles is given by only that of radiative decay, i.e., $\Gamma_{i,L}=\Gamma_{i,\gamma}$ in this paper.  The width for a dipole photon emission is roughly given by
\begin{equation}
\Gamma_{i,\gamma}\sim N_{\rm eff}~\alpha R_{i,\gamma}^2 \Delta E_{\rm r}^3,
\label{eq7}
\end{equation}
where
$\alpha$ is the fine structure constant, and
$\Delta E_{\rm r}$ is a typical interval between energy levels of resonances.  In the above equation, $N_{\rm eff}\sim {\mathcal O}(1)$ is an effective number of final states to which the resonance $i$ can transit by emission of electric dipole photon.

A decay width for an emission of $X_1$ is contributed possibly from multiple partial waves with large penetration probabilities, i.e., $P_{i,j,l}\sim 1$ for $l\lesssim l_{\rm max}$, because of suppressed centrifugal potentials,
\begin{equation}
\Gamma_{i,X}\sim \frac{3 v_{i,X}}{R_{i,X}} \sum_{l} \theta_{i,X,l}^2,
\label{eq8}
\end{equation}
\begin{equation}
v_{i,X}\sim \sqrt[]{\mathstrut \frac{2E}{\mu_{X_1+X_2}}}\approx \sqrt[]{\mathstrut \frac{4E}{m_Y}}
\label{eq9}
\end{equation}
We define the sum of reduced widths as $\theta_{i,X}^2 \equiv \sum_{l} \theta_{i,X,l}^2$.  It should be noted that the conservation of angular momentum limits possible angular momenta $l$ between $X_1$ and $X_2$ to be reached from a resonance of a certain spin number.  We then obtain an equation, i.e.,
\begin{equation}
\Gamma_{i,X}\sim \frac{3}{R_{i,X}}\sqrt[]{\mathstrut \frac{4E}{m_Y}} \theta_{i,X}^2,
\label{eq10}
\end{equation}

Experiments of heavy quarkonia~\cite{PDG2010} show that a typical decay width of such states are $\gtrsim O(10~{\rm MeV})$.  It is assumed that intervals between energy levels of resonances lying above the $X_1+X_2$ threshold [or bound states of $(Y\bar{Y})^\ast$ lying slightly below the threshold] is $\Delta E_{\rm r}\sim 50-100$~MeV \cite{Jacoby:2007nw}.  These intervals are not so large compared with the decay width and the related energy scale, i.e., $T_{\rm C}$.  We assume the equation, i.e., $|E-E_i|\leq \Delta E_{\rm r} \lesssim \Gamma_i$.  The following approximate equation then holds:
\begin{equation}
\sigma\approx \frac{\pi}{2\mu E} \sum_i \omega_i \frac{\Gamma_{i,X} \Gamma_{i,L}}{(\Gamma_i/2)^2}\equiv \frac{\pi}{2\mu E} N_{\rm res}\overline{\left\{\omega \left[\frac{\Gamma_X \Gamma_L}{(\Gamma/2)^2}\right]\right\}},
\label{eq11}
\end{equation}
where
$N_{\rm res}$ is the effective number of resonances through which the reaction proceeds.
In the last equality, we defined $\overline{\left\{\omega \left[\frac{\Gamma_X \Gamma_L}{(\Gamma/2)^2}\right]\right\}}$ as an average quantity per resonance without distinguishing magnetic substates.

We assume that spins of $X_1$ and $X_2$ are zero for simplicity expecting that they are small.  A number of resonances with relative angular momentum $l$ is equal to that of magnetic substates, i.e., $\omega\sim (2l+1)$ approximately.  In addition, we assume that typically one resonance for each value of $l$ works as an intermediate state in the reaction [Eq. (\ref{eq2})].  We then rewrite the expression for the cross section as given by
\begin{eqnarray}
\sigma&\equiv&\frac{\pi (l_{\rm max}+1)^2}{2\mu E}\overline{\left[\frac{\Gamma_X \Gamma_L}{(\Gamma/2)^2}\right]}\nonumber\\
&\sim&0.37f_{\rm w}~{\rm fm}^2\left(\frac{E_{\rm C}}{300~{\rm MeV}}\right)^{-1/2}\left(\frac{m_Y}{1~{\rm TeV}}\right)^{-1}\nonumber\\
&&\times \left(\frac{l_{\rm max}+1}{30}\right)^2 \left(\frac{E}{300~{\rm MeV}}\right)^{-1/2}.\nonumber\\
\label{eq12}
\end{eqnarray}
We defined $\overline{\left[\frac{\Gamma_X \Gamma_L}{(\Gamma/2)^2}\right]}$ as an average quantity per magnetic substate.   The factor, i.e. $f_{\rm w}\equiv \overline{\left[\frac{\Gamma_X \Gamma_L}{(\Gamma/2)^2}\right]}_{E=E_{\rm C}}$, was defined as the value at $E=E_{\rm C}$ (corresponding to $T=T_{\rm C}$).  We used an equation, i.e., $\overline{\left[\frac{\Gamma_X \Gamma_L}{(\Gamma/2)^2}\right]}=f_{\rm w}\sqrt{\mathstrut E/E_{\rm C}}$, where the $E^{1/2}$ scaling derives from the width for an emission of $X$ [Eq. (\ref{eq10})].

This equation indicates that the cross section can be as large as that of normal hadrons if the following two conditions are satisfied: 1. a large number of partial waves ($l_{\rm max}\sim 30$~\cite{Kang:2006yd}) can contribute to the reaction, and 2. decay widths to channels for emission of light particles $L$ are as large as those for $X_1+X_2$ channels.  

The factor, i.e., $f_{\rm w}$ can be approximately unity at maximum.  It can, however, be naturally small if $\Gamma_X>\Gamma_L$ or $\Gamma_X<\Gamma_L$.  The ratio between the decay widths is given [Eqs. (\ref{eq7}) and (\ref{eq10})] by
\begin{eqnarray}
\frac{\Gamma_{i,X}}{\Gamma_{i,L}}&\sim& \frac{3 v_{i,X} \theta_{i,X}^2}{N_{\rm eff}\alpha R_{\rm h}^3\Delta E_{\rm r}^3}\nonumber\\
&\sim&1.4\times 10^4~N_{\rm eff}^{-1} \left(\frac{\theta_{i,X}^2}{1}\right)^{2}\left(\frac{m_Y}{1~{\rm TeV}}\right)^{-1/2}\nonumber\\
&&\times \left(\frac{E}{300~{\rm MeV}}\right)^{1/2} \left(\frac{R_{\rm h}}{1~{\rm GeV}^{-1}}\right)^{-3} \nonumber\\
&&\times \left(\frac{\Delta E_{\rm r}}{100~{\rm MeV}}\right)^{-3}.
\label{eq13}
\end{eqnarray}
In deriving the first line of this equation, we assumed that both of interaction radii at decay of $i$ to $X$ and to $\gamma$ are roughly equal to $R_{\rm h}\sim 1$~GeV$^{-1}$, i.e., the radius of heavy hadrons, i.e., $X_i$.  Unless a condition $\Gamma_X \sim \Gamma_L$ is satisfied, values of the cross section reduces by the factor $f_{\rm w}$ [see Eq. (\ref{eq12})].

\subsection{Effective cross section of $Y\bar{Y}$ annihilation}

At temperature $T\lesssim 200$~MeV, the annihilation proceeds in three stages: a formation of bound states $(Y\bar{Y})$, their transitions to lower energy levels, and the annihilation of $Y\bar{Y}$ pair inside the bound states.  If a bound state tends to be destroyed into separated heavy particles at collisions with thermal particles before processed by $Y\bar{Y}$ annihilation, the state is not an effective path for the annihilation~\cite{Kang:2006yd,Jacoby:2007nw}.  We then introduce an effective cross section of annihilation contributed from only effective paths for annihilation.  It is given by
\begin{eqnarray}
\sigma^{\rm eff}(m_Y, T)&\equiv&\frac{\pi [l_{\rm max}^{\rm eff}(m_Y, T)+1]^2}{2\mu E_{\rm C}^{1/2} E^{1/2}}f_{\rm w}\nonumber\\
&\sim&4.1f_{\rm w}\times 10^{-4}~{\rm fm}^2~\left[l_{\rm max}^{\rm eff}(m_Y, T)+1\right]^2 \nonumber\\
&&\times \left(\frac{E_{\rm C}}{300~{\rm MeV}}\right)^{-1/2} \left(\frac{m_Y}{1~{\rm TeV}}\right)^{-1} \nonumber\\
&&\times \left(\frac{T}{200~{\rm MeV}}\right)^{-1/2},
\label{eq14}
\end{eqnarray}
where $l_{\rm max}^{\rm eff}(m_Y, T)$ is the maximum of angular momenta of partial waves available in effectively annihilating $Y\bar{Y}$.  The maximum angular momentum is given in Sec.~\ref{sec3}.

The rate for annihilation of $Y$ is given by
\begin{eqnarray}
\Gamma_{\rm ann}=\Gamma_{\rm for}\frac{\Gamma_{\rm cas}}{\Gamma_{\rm des}+\Gamma_{\rm cas}}\sim \left\{ \begin{array}{ll}
\Gamma_{\rm for}~~~~~~~~~({\rm if}~\Gamma_{\rm cas}\gg \Gamma_{\rm des})\\
\Gamma_{\rm for}\frac{\Gamma_{\rm cas}}{~\Gamma_{\rm des}}~~({\rm if}~\Gamma_{\rm des}\gg \Gamma_{\rm cas})
\end{array} \right.,\nonumber\\
\label{eq15}
\end{eqnarray}
where
$\Gamma_{\rm for}$ is the rate for bound state $(Y\bar{Y})$ formation, and
$\Gamma_{\rm cas}$ and $\Gamma_{\rm des}$ are the rates of the bound states for cascade down energy levels and for destruction by thermal background particles, respectively.

We make a toy model:  An annihilation cross section for a partial wave $l$ is given by a cross section of bound state formation if $\Gamma_{\rm cas}\geq \Gamma_{\rm des}$, and is zero if $\Gamma_{\rm cas}< \Gamma_{\rm des}$.  The total cross section is then given by Eq. (\ref{eq14}) with $l_{\rm max}^{\rm eff}(m_Y, T)$ changing, like the step function, at temperatures where $\Gamma_{\rm cas}=\Gamma_{\rm des}$ is satisfied for respective angular momenta.

\subsubsection{Cascade and destruction rates (low-lying $l=0$ states)}

The annihilation rate for an $l=0$ bound state is much larger than the destruction rate, i.e., $\Gamma_{\rm cas} \gg \Gamma_{\rm des}$.  Tightly-bound states with small radii have energy eigenvalues and wave functions determined predominantly by the attractive Coulomb-type QCD potential at small radii.  Their structures are, therefore, very similar to that of atoms.  

The annihilation rate for $(Y\bar{Y})_{l=0}\rightarrow q\bar{q}, ...$ is estimated~\cite{Kang:2006yd} similarly to that of positronium.  It is $\Gamma_{\rm ann}\sim \alpha_{\rm QCD}^5\mu=5\times 10^{-3}~{\rm GeV} (\alpha_{\rm QCD}/0.1)^5(m/1~{\rm TeV})$ with $\alpha_{\rm QCD}$ the strong coupling constant.  

The rate for destruction by thermal photons $(Y\bar{Y})_{l=0} +\gamma\rightarrow X_1+X_2$ is estimated using the detailed balance relation~\cite{pagel1997book} as follows:  The cross section for the inverse reaction, i.e., $X_1+X_2 \rightarrow (Y\bar{Y})_{l=0} +\gamma$, through an intermediate state $(Y\bar{Y})^\ast$ with angular momentum $l$ is roughly given by
\begin{equation}
\sigma\sim \frac{\pi\omega f_{\rm w}}{2\mu E_{\rm C}^{1/2} E^{1/2}} \sim \frac{\pi\left(2l+1 \right) f_{\rm w}}{m_Y E_{\rm C}^{1/2} E^{1/2}}.
\label{eq16}
\end{equation}

Supposing the usual atomic model, the number of states $(Y\bar{Y})^\ast$ of angular momentum $\sim l$ is $\sim l_{\rm max}-l+1$, where $l_{\rm max}$ is the maximum angular momentum of all bound states.  There are thus many exit channels generally.  Assuming that all partial cross sections for final states of a given $l$ are equal, the cross section for a fixed final state without distinguishing substates is given by
\begin{equation}
\sigma\sim \frac{\pi\left(2l+1 \right) f_{\rm w}}{(l_{\rm max}-l+1) m_Y E_{\rm C}^{1/2} E^{1/2}}.
\label{eq17}
\end{equation}

The destruction rate is derived using this cross section and the detailed balance relation.  It is then given by
\begin{eqnarray}
\Gamma_{\rm des}&=&n_\gamma \langle \sigma c \rangle\nonumber\\
&\sim&\frac{1}{4\sqrt[]{\mathstrut \pi}} \frac{T^{3/2}}{E_{\rm C}^{1/2}} f_{\rm w} \frac{1}{\left(l_{\rm max}-l+1 \right)} \exp(-E_{\rm th}/T)\nonumber\\
&=&3\times 10^{-11}~{\rm GeV}\left(\frac{f_{\rm w}}{0.01}\right)\left(\frac{E_{\rm C}}{300~{\rm MeV}}\right)^{-1/2}\nonumber\\
&& \times\left(\frac{l_{\rm max}-l+1}{30}\right)^{-1} \left(\frac{T}{200~{\rm MeV}}\right)^{3/2} \nonumber\\
&&\times \left[\frac{\exp(-E_{\rm th}/T)}{4\times 10^{-6}}\right],
\label{eq18}
\end{eqnarray}
where
$n_\gamma$ is the background photon number density,
$\langle \sigma c \rangle$ is the thermal average value of $\sigma c$,
$\sigma(E)$ is the cross section at energy $E$,
$E_{\rm th}=Q$ is the energy threshold of the destruction reaction, and 
$Q\geq 0$ is the reaction $Q$-value.

The binding energy of states with main quantum number $n$ is $E_{\rm B}\sim \mu\alpha_{\rm QCD}/(2n^2)=2.5~{\rm GeV}/n^2(\alpha_{\rm QCD}/0.1)^2(m_Y/1~{\rm TeV})$.  It can be significantly higher than the confinement temperature scale.  In that case, the destruction rate is hindered since only photons energetic enough to dissociate the state can react.  The bound $l=0$ states produced via $X_1+X_2$ collisions thus annihilate before being destroyed by photons.

\subsubsection{Cascade and destruction rates (high-lying $l\geq 1$ states)}

The annihilation of $l\geq 1$ states is hindered because of large angular momenta and low values of wave functions at $r=0$~\footnote{The radial wave function for a system with Coulomb potential scales as $R_{n,l}(r)\propto r^l$.}.  The fate of the states then depends on the rate for transition to states of lower energy and angular momentum, and that for destruction by photons.

Because of color confinement, the predominant bound-bound transition by the strong interaction associates with transitions of at least second order in perturbative color electric or magnetic $2^l$-pole moments (E$l$ or M$l$, respectively)~\cite{Brambilla:2010cs}.  The final state is a bound state, i.e., $(Y\bar{Y})^\ast$, and light hadrons.  Transitions of $\Delta l >1$ to final states with smaller angular momenta are hindered since they need high multipole transitions.

An effective process reducing angular momenta is the dipole photon emission if the $Y$ particle has a finite charge.  The transition rate is given by $\Gamma_{\rm cas}\sim \alpha r_{\rm Bohr}^2 \Delta E^3\sim m_Y \alpha \alpha_{\rm QCD}^4/16$.  An only difference from atomic dipole transition rate originates from the tighter binding by the Coulomb-like QCD force.  The QCD force shrinks wave functions and enhances the transition strength.  

In particle data~\cite{PDG2010}, it is seen that bottomonia (charmonia) bound against decay into $B\bar{B}$ ($D\bar{D}$) channels predominantly decay into bottomonia (charmonia) of lower levels by emitting one photon or two pions except for a decay of $b$ or $\bar{b}$ ($c$ or $\bar{c}$) itself.  The relatively strong transition associated with a two-pion emission between two color-singlet states is via color E1-E1 interaction and the combined angular momentum of pions is $l=0$~\cite{Brambilla:2010cs}.  The most strong transition with a change in angular momentum is, therefore, via a photon emission of mainly $l=1$.  We then assume that the most effective process reducing angular momenta of the system is the dipole photon emission, and consider this process only.

In addition to the spontaneous dipole photon emission, collisional deexcitation reactions can be important \cite{Jacoby:2007nw,Nussinov:2009hc}.  Jacoby \& Nussinov estimated that the cross section for collisional deexcitation is smaller than that for breakup of $(Y\bar{Y})$ into $X_1+X_2$ by $\sim 10^{-2}$--$10^{-3}$.  The deexcitation reaction triggered by a photon, i.e.,
\begin{equation}
(Y\bar{Y})+\gamma \rightarrow (Y\bar{Y})^\prime+\gamma \nonumber\\,
\label{eq19}
\end{equation}
through resonant states is expected to be hindered by a factor $f_{\rm w}^\prime\equiv \overline{\left[\Gamma_\gamma \Gamma_\gamma^\prime/(\Gamma/2)^2\right]}$ [see Eq. (\ref{eq12})].  For example, when the radiative decay width is typically $10^{-2}$ times the total width, the hindrance factor is $f_{\rm w}^\prime\sim 10^{-4}$.  The cross section is then $10^{-2}$ times as large as that for the $(Y\bar{Y})$ breakup [the inverse reaction of Eq. (\ref{eq2})].  

The resonant deexcitation reaction triggered by a pion, i.e.,
\begin{equation}
(Y\bar{Y})+\pi \rightarrow (Y\bar{Y})^\prime+\pi \nonumber\\,
\label{eq20}
\end{equation}
is hindered.  This is because the pion in the final state has $l=2$ relative angular momentum~\cite{Jacoby:2007nw}.  Reactions involving small reduced masses and finite angular momenta are, however, suppressed by a large Centrifugal potential [cf. Eq. (\ref{eq3})].  Thus, the resonant deexcitation cross sections would be smaller than that for the $(Y\bar{Y})$ breakup.  

In this study we assume that radiative resonant reactions have a strength of $f_{\rm w}^\prime=10^{-2}f_{\rm w}$, and are the dominant deexcitation process as one example case.  The deexcitation cross section is, however, uncertain significantly.  We then adopt the rate given [cf. Eq. (\ref{eq18})] by
\begin{eqnarray}
\Gamma_{\rm cas}^{\rm co}&=&\frac{1}{4\sqrt[]{\mathstrut \pi}} \frac{T^{3/2}}{E_{\rm C}^{1/2}} f_{\rm w}^\prime \frac{1}{\left(l_{\rm max}-l+1 \right)} \nonumber\\
&=&8\times 10^{-8}~{\rm GeV}\left(\frac{f_{\rm w}^\prime}{10^{-4}}\right)\left(\frac{E_{\rm C}}{300~{\rm MeV}}\right)^{-1/2}\nonumber\\
&& \times\left(\frac{l_{\rm max}-l+1}{30}\right)^{-1} \left(\frac{T}{200~{\rm MeV}}\right)^{3/2}.
\label{eq21}
\end{eqnarray}

As for destruction processes of $l\geq 1$ bound states, we assume that bound-bound transitions are mainly through the dipole photon absorption which is inverse of the emission process mentioned above.  Bound-free transitions would be through absorptions of photons with $l=1$ or a two-pion pair with $l=0$.  The latter is neglected in the present estimation.  The transition rate is then given by Eq. (\ref{eq18}).

\subsection{Comparison of $(Y\bar{Y})$ formation rate and the Hubble expansion rate}

In this subsection, we describe a way to estimate the relic abundances of SIMP, i.e., $X$, for several possible situations.  In what follows, the MCP, i.e., $Y$, is assumed to be quarklike as an example.  The most important process of bound state formation and its $Q$-value are then
\begin{eqnarray}
&&~~~Y\bar{q}+\bar{Y}q \leftrightarrow (Y\bar{Y})+\{\gamma,~\bar{q}q,~...\},\nonumber\\
&&\hspace{-0.5em}Q=\left\{E_{\rm B}[(Y\bar{Y})]+E_{\rm B}(\gamma,~\bar{q}q,~...)\right\}-[E_{\rm B}(Y\bar{q})+E_{\rm B}(\bar{Y}q)]\nonumber\\
&&\hspace{-0.5em}~~~\sim E_{\rm B}[(Y\bar{Y})]-m_{\{\gamma~{\rm or}~\bar{q}q~{\rm or}~...\}}+2\Delta m(Y\bar{q}),
\label{eq22}
\end{eqnarray}
where
$E_{\rm B}(i)$ is the binding energy of species $i$.  The binding energy is the energy level of (compound) particle measured from the sum of rest masses of constituents.  The gain in mass by a binding can be defined as $\Delta m(i)=-E_{\rm B}(i)$, and we have assumed $\Delta m(Y\bar{q})=\Delta m(\bar{Y}q)$.

From Eq. (\ref{eq14}), the reaction rate is given by
\begin{equation}
\langle \sigma^{\rm eff} v\rangle =\frac{2\pi (l_{\rm max}^{\rm eff}+1)^2 f_{\rm w}}{m_Y^{3/2} E_{\rm C}^{1/2}}.
\label{eq23}
\end{equation}
We note that, if $Q$-values were negative for production of some $(Y\bar{Y})^\ast$ states, then their rates are small since only photons or pions energetic enough to induce the endoergic reaction can play a role.  

Under the assumption of $f_{\rm w}\sim 0.01$ and $l_{\rm max}^{\rm eff}\sim 30$, the reaction rate of bound state formation is given by
\begin{eqnarray}
\Gamma_{\rm for}&=&n_Y\langle \sigma^{\rm eff} v\rangle\nonumber\\
&=&\left[\frac{11}{4} \frac{2\zeta(3)}{\pi^2} T^3 \eta \frac{n_Y}{n_b}\right] \left[\frac{2\pi (l_{\rm max}^{\rm eff}+1)^2 f_{\rm w}}{m_Y^{3/2} E_{\rm C}^{1/2}}\right]\nonumber\\
&=&8\times 10^{-23}~{\rm GeV}~\left(\frac{f_{\rm w}}{0.01}\right) \left(\frac{m_Y}{1~{\rm TeV}}\right)^{-3/2} \nonumber\\
&&\times \left(\frac{E_{\rm C}}{300~{\rm MeV}}\right)^{-1/2} \left(\frac{\eta}{6\times 10^{-10}}\right) \left(\frac{n_Y/n_b}{10^{-6}}\right)\nonumber\\
&&\times \left(\frac{T}{40~{\rm MeV}}\right)^3 \left(\frac{l_{\rm max}^{\rm eff}+1}{30}\right)^2 ,
\label{eq24}
\end{eqnarray}
where
$(11/4)$ is a factor concerning $e^\pm$ annihilation at $T\lesssim m_e/3$ with $m_e$ the mass of electron,
$\zeta(3)=1.2021$ is the third Riemann's zeta function,
$\eta$ is the baryon to photon number ratio of the present Universe,
$n_Y$ and $n_b$ are the number densities of $Y$ and baryon, respectively.

The Hubble expansion rate is given by
\begin{eqnarray}
H&=&\frac{2}{3\sqrt[]{\mathstrut 5}} \frac{\pi^{3/2}}{m_{\rm Pl}}
 g_\ast^{1/2} T^2\nonumber\\
&=&7.1\times 10^{-22}~{\rm GeV}~\left(\frac{g_\ast}{10.75}\right)^{1/2} \left(\frac{T}{40~{\rm MeV}}\right)^2.~~~~
\label{eq25}
\end{eqnarray}

The $Y\bar{Y}$ annihilation does not operate effectively if the formation rate is smaller than the expansion rate, i.e., $H > \Gamma_{\rm for}$.  In general, bound states with smaller binding energies become less subject to destruction by thermal particles at lower temperatures.  In such low temperature environments, however, the bound state formation might be already inefficient because the ratio of rates scales as $\Gamma_{\rm for}/H\propto (l_{\rm max}^{\rm eff}+1)^2T$.

\subsection{Transition of $Y\bar{q}\rightarrow Yqq$}

At $T\sim 40$~GeV, the collisional annihilation of thermal baryons containing quarks with antibaryons is thought to freeze out.  Its number density is then fixed except for the dilution by cosmic expansion.  Abundances of antinucleons keep decreasing via annihilation with abundant nucleons.  This is because there are extra quarks over antiquarks, that is a finite baryon number density in the Universe.  The reaction
\begin{equation}
Y\bar{q}+qqq \leftrightarrow Yqq+\bar{q}q
\label{eq26}
\end{equation}
then drives the $Y\bar{q}$ hadron to $Yqq$~\cite{Nardi1990,DiazCruz:2007fc,Jacoby:2007nw}.  The temperature for operation of this transition is estimated using the Saha equation, i.e.,
\begin{eqnarray}
\frac{n_{Y\bar{q}}{n_{qqq}}}{n_{Yqq}{n_{\bar{q}q}}}&=&\frac{g_{Y\bar{q}} g_{qqq}}{g_{Yqq} g_{\bar{q}q}}
\left(\frac{\mu_{Y\bar{q}+qqq}}{\mu_{Yqq+\bar{q}q}}\right)^{3/2}\exp(-Q/T),~~~~~
\label{eq27}
\end{eqnarray}
where
$n_i$ is the number density of species $i$.

Only pions and nucleons are considered as mesons and baryons, respectively, in a thermal bath.  The equations, $\bar{q}q=\pi$, $J_{\bar{q}q}=0$, $qqq=N$ and $J_{qqq}=1/2$ then hold.  Heavy hadrons, i.e., $Yqq$ and $Y\bar{q}$, are assumed to exist exclusively in only the ground state.   Using the rough assumption that the number abundance of pion is given by the classical Boltzmann distribution, we obtain a ratio between number abundances of $Yqq$ and $Y\bar{q}$:
\begin{eqnarray}
\frac{n_{Yqq}}{n_{Y\bar{q}}}&=&\frac{n_N}{n_\pi} \frac{1}{2} \left(\frac{m_\pi}{m_N}\right)^{3/2}\exp(Q/T)\nonumber\\
&\sim&\frac{11\zeta(3)}{3\sqrt[]{\mathstrut 2\pi}}\eta \left(\frac{T}{m_N}\right)^{3/2}\exp\left[\frac{Q+m_\pi}{T}\right]\nonumber\\
&=&\left(\frac{\eta}{6.2 \times 10^{-10}}\right) \left(\frac{T}{37~{\rm MeV}}\right)^{3/2}\nonumber\\
&&\times \left\{\frac{\exp\left[(Q+m_\pi)/T\right]}{\exp(m_N/37~{\rm MeV})}\right\},
\label{eq28}
\end{eqnarray}
where
$m_N=0.94$~GeV is the nucleon mass,
and $m_\pi=0.14$~GeV is the pion mass.

In this study, we take two different $Q$-values, i.e., $Q=\left[E_{\rm B}(Yqq)+E_{\rm B}(\bar{q}q)\right]-[E_{\rm B}(Y\bar{q})+E_{\rm B}(qqq)]$.  The Saha equation [Eq. (\ref{eq27})] describes that $Y\bar{q}$ particles are processed into $Yqq$ particles at $T_{\rm tra}\lesssim 40$~MeV depending on the $Q$-value.  $T_{\rm tra}$ is the transition temperature below which the condition, i.e., $n_{Yqq}/n_{Y\bar{q}}>1$, is satisfied.  An important point is that the $Q$-value is large compared with the temperature $T$.  If a $Q$-value were smaller, the $Y$ particle is mainly confined in $Y\bar{q}$ until the temperature becomes much smaller (see footnote 1 on page 2 of Ref. \cite{Jacoby:2007nw}).

Since there are only small abundances of antibaryons at this low temperature, the $\bar{Y}$ particles are still confined in $\bar{Y}q$.  They do not experience the corresponding reaction triggered by antibaryons.

\subsection{Annihilation through $Yqq+\bar{Y}q$ reaction}

After the transformation, i.e., $Y\bar{q} \rightarrow Yqq$, the main reaction for bound state formation changes to
\begin{equation}
Yqq+\bar{Y}q \rightarrow (Y\bar{Y})+qqq.
\label{eq29}
\end{equation}
This forward reaction, however, proceeds with a nucleon creation. Therefore, unless energy levels of $Yqq$ and $\bar{Y}q$ are rather high or the binding energy of $(Y\bar{Y})$ is enough large, the inverse reaction tends to dominate because of negative $Q$-value.  

The reaction rate of bound state formation is estimated as follows similarly to the calculation of Eq. (\ref{eq24}).  As the temperature decreases and the number density of energetic $X$ particles decreases, resonant reaction rates for resonances of high energies are expected to become smaller.  Such an inactivation of resonant process is, however, very complicated and not studied in the present study.

\subsubsection{Case of $[\Delta m(Yqq)+\Delta m(\bar{Y}q)]>m_N$}

The rate is given by
\begin{eqnarray}
\Gamma_{\rm for}&=&n_Y\langle \sigma^{\rm eff} v\rangle\nonumber\\
&=&1.8 f_{\rm w} \times 10^{-22}~{\rm GeV}~\left(\frac{m_Y}{1~{\rm TeV}}\right)^{-3/2} \nonumber\\
&&\times \left(\frac{E_{\rm C}}{300~{\rm MeV}}\right)^{-1/2} \left(\frac{\eta}{6\times 10^{-10}}\right) \left(\frac{n_Y/n_b}{1.7\times 10^{-7}}\right)\nonumber\\
&&\times \left(\frac{l_{\rm max}^{\rm eff}+1}{30}\right)^2 \left(\frac{T}{20~{\rm MeV}}\right)^3 ,
\label{eq30}
\end{eqnarray}
where
we roughly assumed that reactions from all initial partial waves contribute to the $(Y\bar{Y})$ formation.  Destructions of $(Y\bar{Y})$ states after their formation by nucleons or photons were, therefore, neglected since the temperature is already rather low.  

Final abundances of $Yqq$ and $\bar{Y}q$ are estimated as values which correspond to the same reaction rates as the Hubble expansion rate.  The ratio between the $Y$ abundance and that of baryon is thus estimated to be $n_Y/n_b\sim2 f_{\rm w}^{-1}\times 10^{-7}$.  The bound state formation after the transformation of $Y\bar{q} \rightarrow Yqq$ is thus somewhat inefficient in reducing the $Y$ abundance.  Relic abundances of exotic hadrons including one $Y$ or $\bar{Y}$ particle then decrease affected by only cosmic expansion since the time of transformation to $Yqq$.

\subsubsection{Case of $[\Delta m(Yqq)+\Delta m(\bar{Y}q)]<m_N$}

The reaction rate is estimated to be smaller than in the previous case because of smaller (or negative) $Q$-values.  We define $E_{\rm th}^{(l)}$ and $\sigma_{\rm h}^{(l)}$ as the threshold energy and the partial cross section, respectively, of reaction [Eq. (\ref{eq29})] whose final state is a $(Y\bar{Y})$ state of angular momentum $l$ and a $qqq$ (nucleon).  The reaction rate is then given by the sum of all angular momenta of $(Y\bar{Y})$ in the final state, i.e., 
\begin{eqnarray}
\Gamma_{\rm for}&\sim&n_Y\sum_l \langle \sigma_{\rm h}^{(l)} v \rangle_{v=\sqrt[]{\mathstrut 2E_{\rm th}^{(l)}/\mu}}^\infty \nonumber\\
&=&n_Y \frac{4\sqrt[]{\mathstrut \pi} f_{\rm w}}{m_Y^{3/2} E_{\rm C}^{1/2}} \sum_l (2l+1) \int_{E_{\rm th}^{(l)}/T}^\infty \sqrt[]{\mathstrut x} \exp(-x)~dx\nonumber\\
&=&n_Y \left[\frac{2\pi l_{\rm min}^2 f_{\rm w}}{m_Y^{3/2} E_{\rm C}^{1/2}}\right.\nonumber\\
&&\left.+\frac{4\sqrt[]{\mathstrut \pi} f_{\rm w}}{m_Y^{3/2} E_{\rm C}^{1/2}} \sum_{l=l_{\rm min}}^{l_{\rm max}} (2l+1) \int_{E_{\rm th}^{(l)}/T}^\infty \sqrt[]{\mathstrut x} \exp(-x)~dx \right]\nonumber\\
&\sim& 7 f_{\rm w} \times 10^{-22}~{\rm GeV}~l_{\rm min}^2 \left(\frac{m_Y}{1~{\rm TeV}}\right)^{-3/2} \nonumber\\
&&\times \left(\frac{E_{\rm C}}{300~{\rm MeV}}\right)^{-1/2} \left(\frac{\eta}{6\times 10^{-10}}\right) \nonumber\\
&&\times \left(\frac{n_Y/n_b}{7\times 10^{-5}}\right) \left(\frac{T}{40~{\rm MeV}}\right)^3 ,
\label{eq31}
\end{eqnarray}
where
$l_{\rm min}$ is the minimum angular momentum of $(Y\bar{Y})$ which have negative $Q$-values, so that reactions producing such states are endoergic.  First and second term in square bracket in the third line correspond to the cross sections for partial waves with $l<l_{\rm min}$ and $l\ge l_{\rm min}$, respectively.  The former component has zero threshold energies so that it does not suffer from Boltzmann suppression factors relating to a high threshold energy.  The second term is neglected in the fourth line since it is generally subdominant to the first term.~\footnote{For example, in Cases 3 and 4 introduced in Sec.~\ref{sec3}, the condition of positive $Q$-value, i.e., $Q\sim E_{\rm B}(Y \overline{Y})-m_N\geq 0$, limits main quantum numbers by $n\lesssim 1.6(\alpha_{\rm QCD}/0.1)(m_Y/1~{\rm TeV})^{1/2}$.  We then take the case of $m_Y=1$~TeV and $\alpha_{\rm QCD}=0.1$.  The minimum angular momentum is then $l_{\rm min}=1$, and the threshold energy is $E_{\rm th}^{(1)}=m_N-E_{\rm B}[(Y\overline{Y})_{n=2}]=0.32$~GeV.  In this case, the second term in Eq. (\ref{eq31}) is less than about a half of the first term.  The values of $l_{\rm min}=1$ (for $m_Y=0.5$ TeV as well as $m_Y=1$ TeV) and $l_{\rm min}=2$ (for $m_Y=3$ TeV) are used in calculations of relic $Y$ abundances in Sec. \ref{sec4}.}

Final abundances of $Yqq$ and $\bar{Y}q$ are estimated to be $n_Y/n_b\sim 7 f_{\rm w}^{-1} l_{\rm min}^{-2} \times 10^{-5}$.  This value is significantly larger than that in the previous case of $[\Delta m(Yqq)+\Delta m(\bar{Y}q)]>m_N$.

\section{Model}\label{sec3}

We describe a model for an estimation of the relic abundance of the exotic heavy hadron $X$.
Table \ref{tab5} lists parameters for four cases we consider in this study.  Case 1 is our standard case.  We chose them in order to clearly show effects of resonance cross section and energy levels of the $X$.

\begin{table}[!t]%[H] add [H] placement to break table across pages
\caption{\label{tab5} Parameters for four cases.}
\begin{ruledtabular}
\begin{tabular}{cccc}
 & $f_{\rm w}$ & $\Delta m(Y\bar{q})$ & $\Delta m(Yqq)$ \\ 
\hline
Case 1 & 0.01 & $0.3~{\rm GeV}$ &  $0.7~{\rm GeV}$ \\
Case 2 & 1 & $0.3~{\rm GeV}$ &  $0.7~{\rm GeV}$ \\
Case 3 & 0.01 & 0 & 0 \\
Case 4 & 1 & 0 & 0 \\
\end{tabular}
\end{ruledtabular}
\end{table}

\subsection{Resonance factor $f_{\rm w}$}

Values for resonance factor $f_{\rm w}$ are adopted as follows.
The particle data on charmonia and bottomonia \cite{PDG2010} provide a possibly useful information.  We consider reactions relating to heavy $c$ or $b$ quarks of the type of Eq. (\ref{eq2}).  The lowest resonance of charmonium above the levels of two-$D$-mesons states ($D^+=c\bar{d}$, $D^0=c\bar{u}$) is $\psi(3770)$.  Its fraction of partial decay width for $D+\bar{D}$ channel relative to the total width is $(93^{+8}_{-9})\times 10^{-2}$, while the sum of those for other $c\bar{c}$ mesons + light particle channels ($\gamma, \pi\pi, \eta, \pi^0$) is $1.5\times 10^{-2}$ \cite{PDG2010}.  The factor of resonance decay widths is then $f_{\rm w}=\Gamma_{\psi(3770),D} \Gamma_{\psi(3770),\{\gamma, \pi\pi, ...\}}/(\Gamma_{\psi(3770)}/2)^2=0.06$.  The rate for the reaction of type [Eq. (\ref{eq2})] through this resonance is, therefore, smaller by the factor $f_{\rm w}$ than the naive estimate of the cross section corresponding to the hadronic size.

The lowest resonance of bottomonium above the levels of two-$B$-mesons states ($B^+=u\bar{b}$, $B^0=d\bar{b}$) is $\Upsilon(4S)$ [or $\Upsilon(10580)$].  Its fraction of partial decay width for $B+\bar{B}$ channel relative to the total width is larger than $0.96$, while the sum of those for other $b\bar{b}$ mesons + light particle channels ($\pi\pi, \eta$) is smaller than $4.4\times 10^{-3}$ \cite{PDG2010}.  The factor is then $f_{\rm w}=\Gamma_{\Upsilon(4S),B} \Gamma_{\Upsilon(4S),\{\pi\pi, \eta ...\}}/(\Gamma_{\Upsilon(4S)}/2)^2<0.018$.

As for reactions of two bound states involving $Y$ particles, resonance widths have large uncertainties.  We assume that the factor is $f_{\rm w}=0.01$ taking account of the value, i.e., $f_{\rm w}<0.018$, derived from particle data on the $b\bar{b}$ resonant state.  This state is composed of the most heavy quark $b$ whose data on resonance decay widths are available.  The factor for $X$ hadrons may be larger or smaller than $f_{\rm w}=0.01$ in fact.  We, therefore, also show results for the case of $f_{\rm w}=1$ as well as those of $f_{\rm w}=0.01$.

\subsection{Energy levels of $Y\bar{q}$ and $Yqq$}

We take two different reaction $Q$-values in order to show that $Q$-values or energy levels of exotic hadrons, i.e., $Y\bar{q}$ and $Yqq$ and so on, are significant in determining final abundance of SIMPs.  In Cases 1 and 2, we take the mass gains of $\Delta m(Y\bar{q})=0.3$~GeV and $\Delta m(Yqq)=0.7$~GeV as assumed in Ref. \cite{Mackeprang:2009ad} in which searches for scatterings of SIMPs ($R$-hadrons) in colliders were studied.  For comparison, in Cases 3 and 4 the mass gains are set to be zero supposing that their amplitudes are not large.

We note that the $Q$-values of the reaction [Eq.~(\ref{eq26})] are $Q=m_N-m_\pi-0.4$~GeV in Cases 1 and 2, and $Q=m_N-m_\pi$ in Cases 3 and 4, respectively.  Such a difference in $Q$-values leads to different transition temperatures [cf. Eq.~(\ref{eq28})], i.e., $T_{\rm tra}\sim 20$~MeV (Cases 1 and 2), and $T_{\rm tra}\sim 37$~MeV (Cases 3 and 4).

\subsection{Effective cross section for $Y\bar{q}+\bar{Y}q$}

We treat the annihilation through the $Y\bar{q}+\bar{Y}q$ reaction in the temperature range of $T\geq T_{\rm tra}$.  The effective cross sections [Eq.~(\ref{eq14})] for Cases 1--4 are given using $l_{\rm max}^{\rm eff}(m_Y,T)$ determined by a procedure described in Sec.~\ref{sec2}B.  Here, each $l_{\rm max}^{\rm eff}(m_Y,T)$ is determined with values of critical temperatures, i.e., $T_{\rm{cri}}(l)$. These critical temperatures are defined by points at which the condition, i.e., $\Gamma_{\rm cas}=\Gamma_{\rm des}$, is satisfied.  Whether the cascade or destruction dominates then changes at $T_{\rm cri}(l)$ for each angular momentum $l$.  A cross section for partial wave $l$ is maximal if $\Gamma_{\rm cas}\geq \Gamma_{\rm des}$, and is zero if $\Gamma_{\rm cas}< \Gamma_{\rm des}$.  

In addition, we adopt an ansatz on the quantity $l_{\rm max}^{\rm eff}(m_Y,T)$ at temperatures below the critical value $T_{\rm cri}(7)$ for Cases 1, 2 and 3.  The ansatz is a linear function of $T$, i.e., 
\begin{equation}
l_{\rm max}^{\rm eff}=7+\frac{T_{\rm cri}(7)-T}{T_{\rm cri}(7)-T_{\rm tra}}\left(\l_{\rm max}-7\right),
\label{eq32}
\end{equation}
where
$l_{\rm max}=30$ is fixed in this study. We calculate critical temperatures $T_{\rm cri}(l)$ below.

\subsubsection{Comparison of rates for cascade and destruction}

The destruction rate is given by Eq. (\ref{eq18}).
The threshold energy is $E_{\rm th}=2.5~{\rm GeV}/n^2(m_Y/1~{\rm TeV})+0.6$~GeV (Cases 1 and 2) and $E_{\rm th}=2.5~{\rm GeV}/n^2(m_Y/1~{\rm TeV})$ (Cases 3 and 4).

The cascade rate is given by sums of spontaneous rates, i.e, $\Gamma_{\rm cas}^{\rm sp}$, [Eqs. (\ref{eq33}) and (\ref{eq34})] and collisional rates, i.e., $\Gamma_{\rm cas}^{\rm co}$, [Eq. (\ref{eq21})].

For the spontaneous cascade rate of states with small $l$, we roughly take the rate for a transition from state $(n,l=n-1)$ to $(n-1,l-1)$ as a typical rate for dipole emission from a state with angular momentum $l$.  It is described by
\begin{eqnarray}
\Gamma_{\rm cas}^{\rm sp}&\sim&\alpha [r_{\rm Bohr}(n-1)]^2 \left[E_{\rm B}(n)-E_{\rm B}(n-1)\right]^3\nonumber\\
&\sim&
%\frac{\alpha \alpha_{\rm QCD}^4m_Y}{2}\left[\frac{(n-1/2)^3}{n^6 (n-1)^2}\right] \nonumber\\
%&=&
\frac{\alpha \alpha_{\rm QCD}^4m_Y}{2}\left[\frac{(l+1/2)^3}{(l+1)^6 l^2}\right],
\label{eq33}
\end{eqnarray}
where
$r_{\rm Bohr}(n)$ and $E_{\rm B}(n)$ are the Bohr radius and the energy eigenvalue of state with main quantum number $n$.

States of $l\sim l_{\rm max}$ with hadronic sizes in space and small binding energies, on the other hand, take longer time to cascade down~\cite{Kang:2006yd,Jacoby:2007nw}.  The rate for an electrically charged $Y$ case is roughly given~\cite{Kang:2006yd} by
\begin{eqnarray}
\Gamma_{\rm cas}^{\rm sp}&\sim&\frac{\alpha \Lambda_{\rm had}^3}{\alpha_{\rm QCD}^{1/2} m_Y^2}\nonumber\\
&\sim&10^{-8}~{\rm GeV}\left(\frac{\Lambda_{\rm had}}{1~{\rm GeV}}\right)^3 \left(\frac{\alpha_{\rm QCD}}{0.1}\right)^{-1/2} \left(\frac{m_Y}{1~{\rm TeV}}\right)^{-2},\nonumber\\
\label{eq34}
\end{eqnarray}
where
$\Lambda_{\rm had}$ is the energy scale of hadronic interaction.

\begin{figure}
\begin{center}
\includegraphics[width=8.0cm,clip]{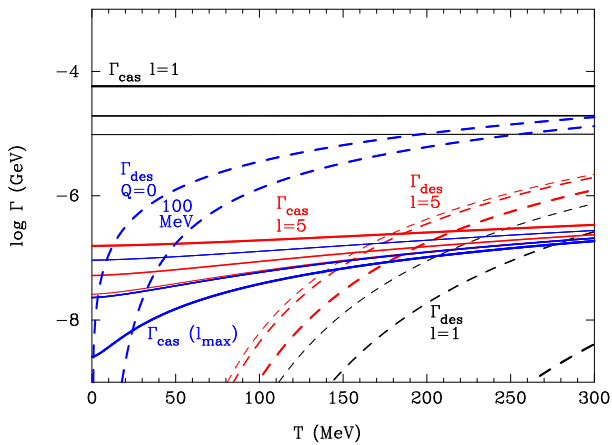}
\caption{Rates for cascading down energy levels, i.e., $\Gamma_{\rm cas}$ (solid lines), and for destruction by thermal photon, i.e., $\Gamma_{\rm des}$ (dashed), of bound states with $l\sim n-1$, as a function of temperature.  The factors of resonance widths are chosen to be $f_{\rm w}=0.01$ and $f_{\rm w}^\prime=10^{-4}$.  The mass gain of $Y\bar{q}$ hadron is $\Delta m(Y\bar{q})=0.3$ GeV.  Masses of the $Y$ particle are taken to be 500 GeV (thin lines), 1 TeV (intermediate) and 3 TeV (thick).  Angular momenta of $(Y\bar{Y})$ are taken to be 1, 5 and $l_{\rm max}$ (from the top line to the bottom for $\Gamma_{\rm cas}$, and the bottom to the top for $\Gamma_{\rm des}$).\label{fig1}}
\end{center}
\end{figure}

Figure \ref{fig1} shows $\Gamma_{\rm cas}$ (solid lines) and $\Gamma_{\rm des}$ (dashed) of bound states with $l\sim n-1$, as a function of temperature.  The colored $Y$ particle has been assumed to have an electric charge.  This figure is for Case 1: The factors of resonance widths are $f_{\rm w}=0.01$ and $f_{\rm w}^\prime=10^{-4}$, and the mass gain of $Y\bar{q}$ hadron is $\Delta m(Y\bar{q})=0.3$ GeV.  Masses of the $Y$ particle are taken to be 500 GeV (thin lines), 1 TeV (intermediate) and 3 TeV (thick).  Angular momenta of $(Y\bar{Y})$ are taken to be 1, 5 and $l_{\rm max}$ (from the top line to the bottom for $\Gamma_{\rm cas}$, and the bottom to the top for $\Gamma_{\rm des}$).

%*******************************************************

%*******************************************************

The $l=1$ excited states produced by the hadronic reaction [Eq. (\ref{eq2})] are found to instantaneously transform to $l=0$ ground state in a temperature range of $T<300$~MeV, as seen from $\Gamma_{\rm cas}\gg \Gamma_{\rm des}$.  The $l=5$ states also tend to transform to $l=4$ state rather than dissociated in a photo-reaction at $T<141-186$.  Critical temperatures change slightly depending upon the mass of the colored particle.  On the other hand, as for $(Y\bar{Y})$ states barely bound relative to the $Y\bar{q}+\bar{Y}q$ separation channel, they can have large angular momenta of $l\lesssim l_{\rm max}$ and small positive reaction $Q$-values of $Q\lesssim 100$~MeV.  Because of the fragility of such barely bound states against background photons, they tend to be destroyed before cascading down to lower energy states until the temperature decreases to $T\sim O(10)$~MeV.

\begin{figure}
\begin{center}
\includegraphics[width=8.0cm,clip]{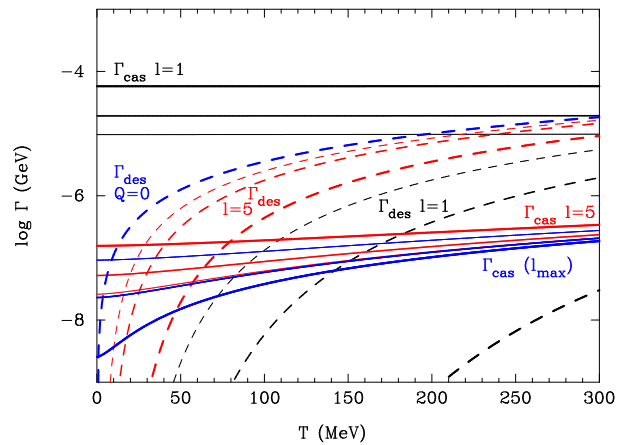}
\caption{Same as in Fig. \ref{fig1} for the case of $\Delta m(Y\bar{q})=0$.\label{fig2}}
\end{center}
\end{figure}

Figure \ref{fig2} shows $\Gamma_{\rm cas}$ (solid lines) and $\Gamma_{\rm des}$ (dashed) similar to Fig. \ref{fig1} for Case 3: The factors of resonance widths are $f_{\rm w}=0.01$ and $f_{\rm w}^\prime=10^{-4}$, and the mass gain of $Y\bar{q}$ hadron is $\Delta m(Y\bar{q})=0$.  The smaller value of $\Delta m(Y\bar{q})$ leads to smaller $Q$-values for respective bound states.  The bound states are, therefore, more easily destroyed by background photons.  Destruction rates are thus generally larger than in Case 1.  The $l=1$ excited states instantaneously transform to the $l=0$ ground state similarly to Case 1.  The $l=5$ states tend to experience transitions to lower states at $T<11-54$.  Barely bound $(Y\bar{Y})$ states tend to be dissociated by background photons until the temperature decreases to $T\sim O(10)$~MeV.

%*******************************************************

%*******************************************************

\subsubsection{Estimation of effective cross sections}

Table \ref{tab1} shows critical temperatures in Case 1.  The first column shows the mass of the $Y$ particle.  The second to ninth columns list critical temperatures for respective $l$ values.  Energy levels naively based on the Coulomb models were assumed.  States with main quantum numbers $n\gtrsim 8$ would be mainly affected by not a Coulomb-type QCD potential but a linear potential.  We do not treat details on loosely-bound energy states originating from this linear potential.  The energy levels and wave functions, however, should be studied for various likely potential models in future to calculate formation cross sections and rates of transitions to lower energy states precisely.

\begin{table}[!t]%[H] add [H] placement to break table across pages
\caption{\label{tab1} Critical temperature of annihilation for Case 1.}
\begin{ruledtabular}
\begin{tabular}{c|cccccccc}
$m_Y$~(TeV) & \multicolumn{8}{c}{$T_{\rm cri}$~(MeV)}\\ 
\cline{2-9}
& $l=0$ & $1$ & $2$ & $3$ & $4$ & $5$ & $6$ & $7$ \\
\hline
0.5 &  608 & 285 & 199 & 165 & 149 & 141 & 137 & 135 \\
1   &  887 & 375 & 243 & 190 & 164 & 151 & 144 & 140 \\
3   & 1820 & 670 & 384 & 271 & 216 & 186 & 169 & 158 \\
\end{tabular}
\end{ruledtabular}
\end{table}

Table \ref{tab2} shows critical temperatures in Case 2 similar to Table \ref{tab1}.  In Case 2, the setup is the same as in Case 1 except for the difference in the resonance factor.  The factor in Case 2, i.e., $f_{\rm w}=1$, is 100 times larger than that in Case 1.  Destruction rates $\Gamma_{\rm des}$ in Case 2 are then 100 times as large as values in Fig. \ref{fig1}.  Because of relatively larger destruction rates, critical temperatures change to lower values.

\begin{table}[!t]%[H] add [H] placement to break table across pages
\caption{\label{tab2} Critical temperature of annihilation for Case 2.}
\begin{ruledtabular}
\begin{tabular}{c|cccccccc}
$m_Y$~(TeV) & \multicolumn{8}{c}{$T_{\rm cri}$~(MeV)}\\ 
\cline{2-9}
& $l=0$ & $1$ & $2$ & $3$ & $4$ & $5$ & $6$ & $7$ \\
\hline
0.5 & 231 & 165 & 148 & 142 & 138 & 136 & 135 & 134 \\
1   & 318 & 199 & 166 & 153 & 146 & 141 & 139 & 137 \\
3   & 646 & 331 & 237 & 197 & 176 & 164 & 156 & 150 \\
\end{tabular}
\end{ruledtabular}
\end{table}

Table \ref{tab3} shows critical temperatures in Case 3.  Because of reaction $Q$-values smaller than in Case 1, the destruction process by background photons is stronger.  This leads to delayed transition, or smaller critical temperatures than in Table \ref{tab1}.

\begin{table}[!t]%[H] add [H] placement to break table across pages
\caption{\label{tab3} Critical temperature of annihilation for Case 3.}
\begin{ruledtabular}
\begin{tabular}{c|cccccccc}
$m_Y$~(TeV) & \multicolumn{8}{c}{$T_{\rm cri}$~(MeV)}\\ 
\cline{2-9}
& $l=0$ & $1$ & $2$ & $3$ & $4$ & $5$ & $6$ & $7$ \\
\hline
0.5 &  381 & 110 &  49 &  27 & 17 & 11 &  8 &  6 \\
1   &  659 & 195 &  88 &  48 & 30 & 20 & 15 & 11 \\
3   & 1599 & 490 & 226 & 126 & 79 & 54 & 39 & 29 \\
\end{tabular}
\end{ruledtabular}
\end{table}

Table \ref{tab4} shows critical temperatures in Case 4.  Because of reaction $Q$-values smaller than in Case 2, the destruction process is stronger.  The transition is then delayed and critical temperatures are smaller than in Table \ref{tab2}.

\begin{table}[!t]%[H] add [H] placement to break table across pages
\caption{\label{tab4} Critical temperature of annihilation for Case 4.}
\begin{ruledtabular}
\begin{tabular}{c|cccccccc}
$m_Y$~(TeV) & \multicolumn{8}{c}{$T_{\rm cri}$~(MeV)}\\ 
\cline{2-9}
& $l=0$ & $1$ & $2$ & $3$ & $4$ & $5$ & $6$ & $7$ \\
\hline
0.5 &  99 &  37 &  19 & 11 &  8 &  6 &  4 &  3 \\
1   & 187 &  71 &  38 & 23 & 16 & 12 &  9 &  7 \\
3   & 516 & 203 & 109 & 68 & 47 & 34 & 26 & 20 \\
\end{tabular}
\end{ruledtabular}
\end{table}

If the colored $Y$ particle were neutral, its rate for cascade down would be small and the transition would be via emissions of light mesons.  Formation cross sections of $(Y\bar{Y})$ bound states [Eq. (\ref{eq12})] are also small since channels of dipole photon emissions are not open in resonant reactions.  In addition, a dissociation rate of the $(Y\bar{Y})$ bound states is small since dissociations triggered by background photons are weak as in the inverse reaction.  Small rates of the  $(Y\bar{Y})$ bound state formation in the case of neutral $Y$ particles would then result in more early freeze-out of annihilation in the Universe.  Final abundances of exotic heavy hadrons are resultingly larger than in the charged case.

\section{Result}\label{sec4}

Relic abundances of exotic heavy hadrons are calculated using the annihilation rates described in Sec.~\ref{sec3}.
We denote hadrons which include one $Y$ particle by $X$, and hadrons which include one $\bar{Y}$ particle by $\bar{X}$.  It should be noted that $\bar{Y}$ particles are not always antiparticles of $Y$s.  The time evolution of the number density of heavy hadron $X$ ($\bar{X}$) is described by the following integrated Boltzmann equation \cite{Gondolo1991}:
\begin{equation}
\frac{dY_{X}}{dt}=\frac{dY_{\bar{X}}}{dt}=-s \langle \sigma (m_Y, T) v \rangle Y_X Y_{\bar X},
\label{eq:Boltzmann}
\end{equation}
where $Y_i = n_i/s$ was defined as a measure of the number of species $i=X$ and $\bar{X}$ per comoving volume \cite{kolb1990}.
$n_i$ is the number density of $i$, 
and $s$ is the entropy density of the Universe.
$\langle \sigma (m_Y, T) v \rangle$ is the effective reaction rate for $Y \bar{Y}$ annihilation via formation of bound states of MCPs $(Y\bar{Y})$ by two SIMPs.
In this equation, the inverse reaction of annihilation, i.e., the heavy hadron production, has been neglected.  Since the temperature of the Universe $T\lesssim 200$~MeV is much lower than the mass of MCP $m\sim O(1~{\rm TeV})$, such a production is never operative.

Effects of the destruction and deexcitation of bound states $(Y\bar{Y})$ by thermal particles, and the transition of heavy hadron species, i.e., $Y\bar{q}\rightarrow Yqq$, are taken into account in the reaction rate $\langle \sigma (m_Y, T) v \rangle$.  We assume that $X$ and $\bar{X}$ have the same number density at the temperature of the QCD phase transition $T_{\rm C} \sim 200$~MeV \footnote{The possibility of asymmetry in the baryon number related to the exotic hadron $X$ has been suggested \cite{Dover:1979sn,Nardi1990}.}.  Then, the equations for time evolutions of $Y_X$ and $Y_{\bar{X}}$ are the same, and the equation, i.e., $Y_{\bar{X}}=Y_{X}$, holds.  The equation is described by
\begin{equation}
\frac{dY_{X}}{dt} = - s \langle \sigma (m_Y, T) v \rangle Y_X^2.
\end{equation}
This equation is solved as a function of temperature $T$ \cite{Gondolo1991}:
\begin{eqnarray}
Y_X (T)&\hspace{-0.5em}=&\hspace{-0.5em}Y_X (T_{\rm C})\left[1 + Y_X (T_C) m_{\rm pl} \sqrt{\frac{\pi}{45}} \right.\nonumber\\
&&\hspace{4.5em}\left. \times \int_{T}^{T_{\rm C}} dT~ \sqrt[]{\mathstrut g_{\rm eff}(T)} \langle \sigma(m_Y, T) v \rangle\right]^{-1}\hspace{-1.em},~~~~~
\end{eqnarray}
where $\sqrt{g_{\text{eff}}(T)} = \sqrt{g_{*}(T)} \left( 1 + \frac{T}{3 g_{*} (T)} \frac{\mathrm{d} g_{*} (T)}{\mathrm{d} T} \right)$ was defined.

\begin{figure}
\begin{center}
\includegraphics[width=8.0cm,clip]{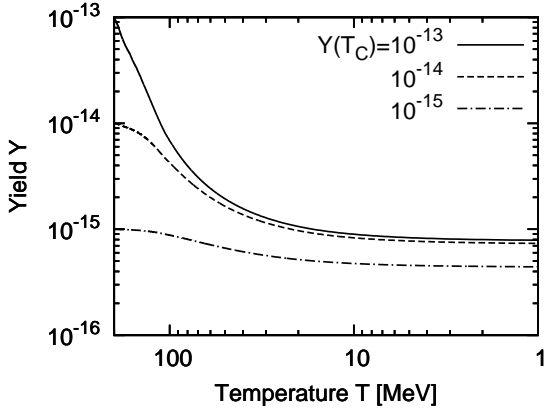}
\caption{Yields of heavy hadrons $X$ and $\bar{X}$ as a function of temperature in the region $T=200$~MeV--$1$~MeV for Case 1.  The mass of the $Y$ is fixed to be $m_Y=1$ TeV.  The solid, dashed and dot-dashed lines correspond to the initial yields of heavy hadrons of $Y (T_C) = 10^{-13}, 10^{-14}$ and $10^{-15}$, respectively.}
\label{fig:case1}
\end{center}
\end{figure}

\begin{figure}
\begin{center}
\includegraphics[width=8.0cm,clip]{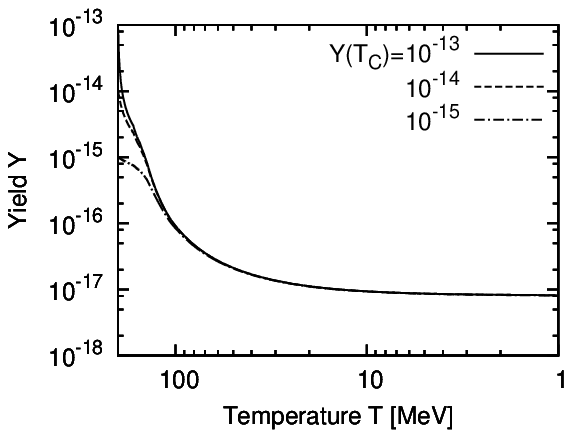}
\caption{Same as in Fig.~\ref{fig:case1} for Case 2 for $m_Y=1$ TeV.}
\label{fig:case2}
\end{center}
\end{figure}

\begin{figure}
\begin{center}
\includegraphics[width=8.0cm,clip]{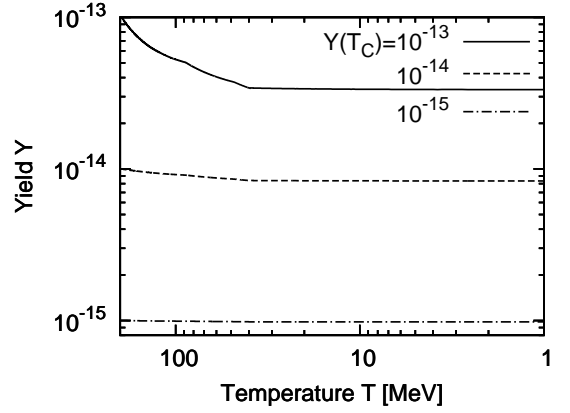}
\caption{Same as in Fig.~\ref{fig:case1} for Case 3 for $m_Y=1$ TeV.}
\label{fig:case3}
\end{center}
\end{figure}

\begin{figure}
\begin{center}
\includegraphics[width=8.0cm,clip]{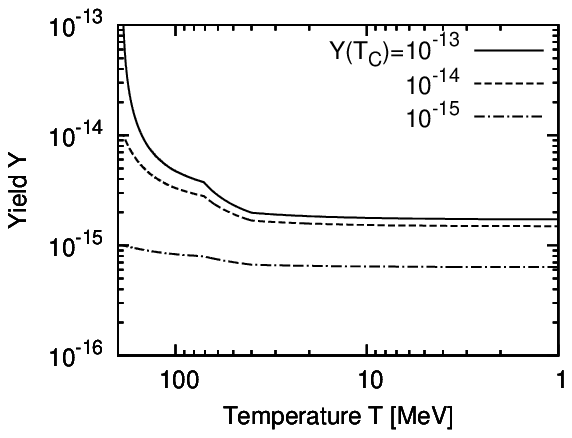}
\caption{Same as in Fig.~\ref{fig:case1} for Case 4 for $m_Y=1$ TeV.}
\label{fig:case4}
\end{center}
\end{figure}

In Figs.~\ref{fig:case1}--\ref{fig:case4}, we plot calculated abundances of $X$ and $\bar{X}$, i.e.,  $Y(T) = Y_X(T) + Y_{\bar{X}} (T) = 2 Y_{X} (T)$ for Cases 1--4 (see Table~\ref{tab5}), respectively.
We took three initial abundances of $Y (T_C) = 10^{-13}, 10^{-14}$ and $10^{-15}$, and results are shown by the solid, dashed, and dot-dashed lines, respectively.
Fig.~\ref{fig:case1} shows results for our standard case, i.e., Case 1.
%%%%%%%FIGURE%%%%%%%%%%%%%%%%%%%%%%
%%%%%%%%%%%%%%%%%%%%%%%%%%%%%%%%%

In Fig.~\ref{fig:case2}, the $f_{\rm w}$ value is 100 times lager than that in Case 1, and the mass gains of $Y \bar{q}$ and $Yqq$ are the same as in Case 1.
This difference results in relic abundances smaller than in Fig.~\ref{fig:case1} by a factor of $\sim$100.
In Fig.~\ref{fig:case2}, the final abundances of exotic hadrons do not depend at all on the initial values of $Y(T_{\rm C})$.
%%%%%%%FIGURE%%%%%%%%%%%%%%%%%%%%%%
%%%%%%%%%%%%%%%%%%%%%%%%%%%%%%%%%

In Fig.~\ref{fig:case3}, while the $f_{\rm w}$ value is the same as in Fig.~\ref{fig:case1}, the mass gains of exotic hadrons are assumed to be zero.
Because of the zero mass gains, the bound state formation rates are smaller and the resulting relic abundances are larger than those in Case 1.
The $Y\bar{Y}$ annihilation at the hadronization of the MCPs $Y$ and $\bar{Y}$ is not so important to the final abundances in Case 3.
%%%%%%%FIGURE%%%%%%%%%%%%%%%%%%%%%%
%%%%%%%%%%%%%%%%%%%%%%%%%%%%%%%%%

Finally, Fig.~\ref{fig:case4} shows results for Case 4 in which the $f_{\rm w}$ value is 100 times lager than in Case 1 and the mass gains are assumed to be zero.
The final abundances in Case 4 are similar to those in Case 1.  This is because the effects of the resonance factor and the mass gains work in different directions and they get balanced out.
%%%%%%%FIGURE%%%%%%%%%%%%%%%%%%%%%%
%%%%%%%%%%%%%%%%%%%%%%%%%%%%%%%%%

Nonsmooth behaviors can be observed in Figs.~\ref{fig:case3} and Fig.~\ref{fig:case4}.  They come from discontinuities in adopted $l_{\text{max}} (m_Y, T)$ values which lead to discontinuities in the effective cross sections [Eq. (\ref{eq14})] and reaction rates.  The discontinuity would be an artificial fake made in our simplified toy model, and not be physical.  We expect that more smooth cross sections would realize in more dedicated models.

In Fig.~\ref{fig:case4}, plateau yields are seen at $T \sim 200$~MeV (leftmost part).  This flat structure results from $\langle \sigma (m_Y, T) v \rangle = 0$.  It reflects that rates for cascading down of $(Y \bar{Y})$ states are smaller than those for their destruction by thermal particles for all partial waves at the high temperature. 

\begin{figure}
\begin{center}
\includegraphics[width=8.0cm,clip]{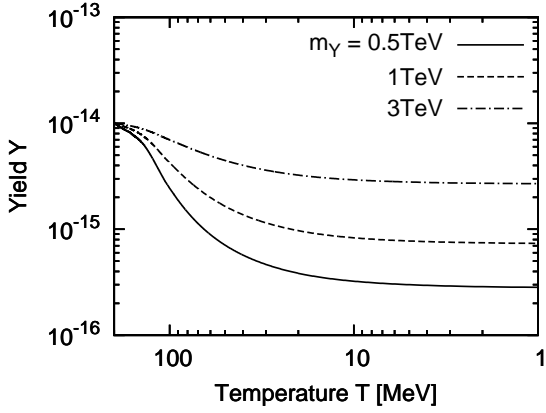}
\caption{Yields of heavy hadrons $X$ and $\bar{X}$ as a function of temperature in the region $T=200$~MeV--$1$~MeV for Case 1.  The initial yield is fixed to be $Y(T_{\rm C})=10^{-14}$.  The solid, dashed and dot-dashed lines correspond to the masses of $Y$ of $m_Y=0.5$, $1$ and $3$ TeV, respectively.}
\label{fig:case1hikaku}
\end{center}
\end{figure}

For illustration of the mass dependence of the relic abundance, in Fig.~\ref{fig:case1hikaku} we plot yields $Y (T)$ for cases of different masses.  The initial yield is fixed to be $Y(T_{\rm C})=10^{-14}$.  The solid, dashed, and dot-dashed lines correspond to the masses of $Y$ of $m_Y=0.5$, $1$ and $3$ TeV, respectively.  One can see that the larger mass results in the larger relic abundance.
This behavior derives from a dependence of the formation rate on mass $m_Y$ by the negative power [cf. Eq. (\ref{eq23})].

%%%%%%%FIGURE%%%%%%%%%%%%%%%%%%%%%%
%%%%%%%%%%%%%%%%%%%%%%%%%%%%%%%%%

In Table~\ref{tab6}, we summarize calculated number ratios of heavy hadrons to normal baryons, i.e., $n_{X+\bar{X}}/n_b$ at $T=1$~MeV.  For Cases 1--4, results are shown for different masses and initial yields.

%%%%%%%%%%TABLE%%%%%%%%%%%%%%%%%%%%%%%%%%%%%
\begin{table}[!t]%[H]
\label{tab6}
\caption{Calculated number ratios of heavy hadrons to normal baryons $n_{X+\bar{X}}/n_b$ at $T=1$~MeV for four Cases (see Table \ref{tab5}) and different masses $m_Y$ and initial yields of heavy hadrons, i.e., $Y(T_{\rm C})$.}
\begin{ruledtabular}
\begin{tabular}{c|c|ccc}
Case   & $Y(T_{\rm C})$         &   $m_Y=0.5$~TeV         &           $1$~TeV       &$3$~TeV \\ 
\hline
       & $10^{-13}$             &$3.4 \times 10^{-6}$     &$9.2 \times 10^{-6}$     &$4.2 \times 10^{-5}$ \\
Case 1 & $10^{-14}$             &$3.3 \times 10^{-6}$     &$8.6 \times 10^{-6}$     &$3.2 \times 10^{-5}$ \\
       & $10^{-15}$             &$2.6 \times 10^{-6}$     &$5.2 \times 10^{-6}$     &$9.2 \times 10^{-6}$ \\
\hline
       & $10^{-13}$             &$3.5 \times 10^{-8}$     &$9.6 \times 10^{-8}$     &$4.5 \times 10^{-7}$ \\ 
Case 2 & $10^{-14}$             &$3.5 \times 10^{-8}$     &$9.6 \times 10^{-8}$     &$4.5 \times 10^{-7}$ \\
       & $10^{-15}$             &$3.5 \times 10^{-8}$     &$9.5 \times 10^{-8}$     &$4.4 \times 10^{-7}$ \\
\hline
       & $10^{-13}$             &$3.3 \times 10^{-4}$     &$3.9 \times 10^{-4}$     &$5.7 \times 10^{-4}$ \\
Case 3 & $10^{-14}$             &$9.3 \times 10^{-5}$     &$9.8 \times 10^{-5}$     &$1.1 \times 10^{-4}$ \\
       & $10^{-15}$             &$1.1 \times 10^{-5}$     &$1.2 \times 10^{-5}$     &$1.2 \times 10^{-5}$ \\
\hline
       & $10^{-13}$             &$2.2 \times 10^{-5}$     &$2.0 \times 10^{-5}$     &$2.1 \times 10^{-5}$ \\
Case 4 & $10^{-14}$             &$1.9 \times 10^{-5}$     &$1.8 \times 10^{-5}$     &$1.8 \times 10^{-5}$ \\
       & $10^{-15}$             &$7.7 \times 10^{-6}$     &$7.5 \times 10^{-6}$     &$7.6 \times 10^{-6}$ \\
\end{tabular}
\end{ruledtabular}
\end{table}
%%%%%%%%%%%%%%%%%%%%%%%%%%%%%%%%%%%%%%%%%%%

\section{Effect of $(YYq)$ heavy hadron}\label{sec5}

We note that states involving two exotic heavy colored particles, i.e., $(YYq)$, would naturally exist if the $YY$ annihilation does not operates inside the hadron.  On the other hand, if two $Y$ particles interact inside the hadron and transform into SM particles, the particles $(YYq)$ play a role in reducing the abundance of $Y$.  Productions or detections of such hadrons with double $Y$s in colliders might be rather difficult.  They are, however, thought to have been produced and also destroyed in a hot epoch of the early Universe.

Although we neglect these states in this study, we comment on their possible effects assuming that they are long-lived.  The most important thing regarding the effects would be reaction $Q$-values which stem from energy levels of relevant hadrons.  Possible processes for production of $(YYq)$ [and $(\bar{Y}\bar{Y}\bar{q})$] are the following three (and similar three for antihadron):

\subsection{$Y\bar{q}+Y\bar{q}$~~~~($T\gtrsim T_{\rm tra}$)}
\begin{eqnarray}
&&~~~Y\bar{q}+Y\bar{q} \leftrightarrow (YYq)+\bar{q}\bar{q}\bar{q},\nonumber\\
&&Q=E_{\rm B}[(YYq)]-m_N+2\Delta m(Y\bar{q}).
\label{eq35}
\end{eqnarray}
In Cases 1 and 2, the $Q$-value is given by $Q\sim E_{\rm B}(YYq)-m_N+0.6~{\rm GeV}$.  Because of the height of the resonance level of $Y\bar{q}$, and the mass of nucleon in the exit channel, the $Q$-value is expected to be negative.  This reaction is then not likely an available path for two $Y$ (or $\bar{Y}$) particles to be bound.  If a $Q$-value is positive and relatively large, e.g. of ${\mathcal O}(100~{\rm MeV})$, however, the colored heavy particles are effectively processed through this reaction, and locked in the $(YYq)$ hadron.

The $(YYq)$ hadron thus produced can be processed through two reactions, i.e.,
\begin{eqnarray}
&&~~~(YYq)+Y\bar{q} \leftrightarrow (YYY)+\{\gamma,~q\bar{q},~...\},\nonumber\\
&&\hspace{-1.em}Q=E_{\rm B}[(YYY)]-m_{\{\gamma,~q\bar{q},...\}}-\{E_{\rm B}[(YYq)]-\Delta m(Y\bar{q})],\nonumber\\
\label{eq36}
\end{eqnarray}
and
\begin{eqnarray}
&&~~~(YYq)+\bar{Y}q \leftrightarrow (Y\bar{Y})+Yqq,\nonumber\\
&&\hspace{-1.em}Q=\{E_{\rm B}[(Y\bar{Y})]-\Delta m(Yqq)\}-\{E_{\rm B}(YYq)-\Delta m[(\bar{Y}q)]\}.\nonumber\\
\label{eq37}
\end{eqnarray}
Whether these two effectively operate depends on their reaction $Q$-values.  However, both reactions possibly have positive $Q$-values.  The colored particle $Y$ would then be predominantly included into $(YYY)$ which is long-lived and $(Y\bar{Y})$ which eventually annihilates.  

The $(YYY)$ hadron is expected to be rather tightly bound because of its large mass.  The only possible reaction processing low-lying states of $(YYY)$ with positive $Q$-value is then
\begin{eqnarray}
&&~~~(YYY)+\bar{Y}q \leftrightarrow (Y\bar{Y})+(YYq),\nonumber\\
&&\hspace{-1.em}Q=\{E_{\rm B}[(Y\bar{Y})]+E_{\rm B}[(YYq)]\}-\{E_{\rm B}[(YYY)]-\Delta m(\bar{Y}q)\}.\nonumber \\
\label{eq38}
\end{eqnarray}
Energy levels of the entrance channel including $(YYY)$ are typically lower than those of the exit channel including $(Y\bar{Y})$, although somewhat excited $(YYY)$ states may be transformed into tightly-bound low-energy levels of $(Y\bar{Y})$.  Then, a significant fraction of initially existent $Y$ might be confined in $(YYY)$ particles.   The $(YYY)$ particle would escape from reducing its abundance since the annihilation $(YYY)$+$(\bar{Y}\bar{Y}\bar{Y})$ is hindered by somewhat larger mass and significantly smaller size of the targets composed of three heavy $Y$s than those composed of a single $Y$ and one light $\bar{q}$ or two light $q$s.  

Final abundances of $(YYY)$ and $(\bar{Y}\bar{Y}\bar{Y})$ in this case would be roughly of the order of initial $Y$ abundance fixed at freeze-out of perturbative annihilation, i.e., $3n_{(YYY)}/n_b\lesssim 10^{-14}$.  Depending on situations, final abundances of $(YYq)$ can be high, i.e., $2n_{(YYq)}/n_b\lesssim 10^{-14}$.

\subsection{$Yqq+Y\bar{q}$~~~~($T\sim T_{\rm tra}$)}

\begin{eqnarray}
&&Yqq+Y\bar{q} \leftrightarrow (YYq)+\{\gamma,~q\bar{q},~...\},\nonumber\\
&&\hspace{-1.em}Q=E_{\rm B}[(YYq)]-m_{\{\gamma,~q\bar{q},~...\}}+[\Delta m(Yqq)+\Delta m(Y\bar{q})].\nonumber\\
\label{eq39}
\end{eqnarray}
In Cases 1 and 2, the $Q$-value is given by $Q\lesssim E_{\rm B}[(YYq)]+1~{\rm GeV}$ depending upon light particles in the final state.  The $Q$-value might be positive and large.  In our estimation, abundances of $Yqq$ and $Y\bar{q}$ are comparable around the transition temperature $T_{\rm tra}\sim 20$~MeV.  At this temperature, this reaction can operate.  If the $Y\bar{Y}$ annihilation through $(YYq)$+$\bar{Y}q$ or $(YYY)$+$\bar{Y}q$ is hindered by possible large negative $Q$-values for the reactions, the $\bar{Y}$ particles would survive being dominantly locked in the $\bar{Y}q$ hadron.  If the reaction [Eq. (\ref{eq39})] effectively operates, final abundances of $(YYq)$ or $(YYY)$, and $\bar{Y}q$ would be roughly of the order of $Y\bar{q}$ abundance at the transition temperature or less, i.e., $3n_{(YYY)}/n_b\lesssim 10^{-8}$--$10^{-4}$ or $2n_{(YYq)}/n_b\lesssim 10^{-8}$--$10^{-4}$, and $n_{\bar{Y}q}/n_b\lesssim 10^{-8}$--$10^{-4}$.  

\subsection{$Yqq+Yqq$~~~~($T\lesssim T_{\rm tra}$)}

\begin{eqnarray}
&&~~~Yqq+Yqq \leftrightarrow (YYq)+qqq,\nonumber\\
&&\hspace{-1.em}Q=E_{\rm B}[(YYq)]-m_N+2\Delta m(Yqq).
\label{eq40}
\end{eqnarray}

In Cases 1 and 2, the $Q$-value is given by $Q\sim E_{\rm B}[(YYq)]-m_N+1.4~{\rm GeV}$.    The $Q$-value is then positive and large more likely than that of the first reaction [Eq. (\ref{eq35})], and less likely than that of the second [Eq. (\ref{eq39})].

If we refer to values of charmed baryon case, the $Q$-value is negative:  The mass of $\Lambda_c^+=udc$ is 2284~MeV.  The mass of $\Xi_{cc}^+=ucc$ is 3519 MeV.  The mass of charm quark is 1.27 GeV.  This mass difference leads to $E_{\rm B}[(YYq)]+\Delta m(Yqq)=-0.979~{\rm GeV}+1.01~{\rm GeV}=-0.03$~GeV \cite{PDG2010}.  Adopting this and the assumption, $\Delta m(Yqq)=0.7$, we derive $Q=-0.27$~GeV.

If the previous two reactions [Eqs. (\ref{eq35}) and (\ref{eq39})] are not operative in producing the $(YYq)$ and this reaction [Eq. (\ref{eq40})] is operative for some reason, final abundances of $(YYq)$ or $(YYY)$ and $\bar{Y}q$ would be roughly of the order of $Y\bar{q}$ abundance at the transition temperature or less, i.e., $3n_{(YYY)}/n_b\lesssim 10^{-8}$--$10^{-4}$ or $2n_{(YYq)}/n_b\lesssim 10^{-8}$--$10^{-4}$, and $n_{\bar{Y}q}/n_b\lesssim 10^{-8}$--$10^{-4}$. 

\section{Conclusions and discussion}\label{sec6}

We have investigated effects of decay properties of resonances made of two long-lived massive colored particles (MCPs), i.e., $(Y\bar{Y})$, and binding energies or energy levels of exotic strongly interacting massive particles (SIMPs) including one MCP in them, on the annihilation of MCPs at color confinement temperature $T_{\rm C}\sim 200$~MeV.  Magnitudes of cross sections of two-SIMP collision for final states including a $(Y\bar{Y})$ is expressed by a parameter $f_{\rm w}$.  In addition, binding energies of SIMPs are taken as other parameters.  We studied effects of the parameters on the $Y\bar{Y}$ annihilation, and resulting relic abundances of SIMPs by calculating with different parameter sets.   

It is assumed that the $(Y\bar{Y})$ bound state formation from initial states of two SIMPs dominantly proceeds through resonances (or barely-bound states) of $(Y\bar{Y})$.  The cross section for bound state formation is then formulated utilizing a well-known expression for resonant cross section.  We suppose that bound states or resonances of MCPs decay into more bound states mainly through the dipole photon emission rather than hadron emissions, based upon an ansatz from particle data on heavy quarkonia \cite{PDG2010}.  In addition, we roughly estimate fates of $(Y\bar{Y})$ states from a comparison of the rate for cascading down energy levels by the spontaneous emission and collisions with background photons to that for being destroyed by collisions.  We then make a toy model of the effective cross section for $Y\bar{Y}$ annihilation by two-SIMP collisions which accounts for only the reaction which ends with the $Y\bar{Y}$ annihilation without dissociated into two SIMPs by background photons.  In this model, we roughly take into account the transition of SIMP, i.e, $Y\bar{q}\rightarrow Yqq$, where $q$ is a light quark.

Evolutions of SIMP abundances are calculated as a function of the temperature for four cases.  Different parameter sets for three initial abundances and three masses of MCP are considered in the calculation.  In Figs.~\ref{fig:case1}--\ref{fig:case4} we show results of abundance evolution at temperature $T\lesssim T_{\rm C}$ for some cases.  Final abundances for all cases are listed in Table \ref{tab6}.   We show that larger $f_{\rm w}$ values leads to smaller final abundances of SIMP, while smaller mass gains of SIMPs lead to larger final abundances.  We then suggest that informations on exotic heavy hadrons such as their energy levels and partial decay widths are necessary to estimate relic abundances of SIMPs precisely.  The calculated relic abundances (Table \ref{tab6}) for respective SIMP species are $2\times 10^{-8}$--$3\times 10^{-4}$ times that of baryon depending on parameters.  This large uncertainty stems from uncertainties in level structures and decay widths of SIMPs, and the mass and the initial abundance at color confinement of MCP.  We finally comment on possible scenarios in a case that long-lived heavy hadrons composed of two $Y$ particles exist.

We note that the relatively small final abundances of SIMPs with a huge uncertainty can be important since even such small amounts of SIMPs possibly leave observational signatures on primordial light element abundances through bindings of SIMPs to nuclei and nuclear reactions of the bound states \cite{Kusakabe:2009jt,Kawasaki:2010yh}.  If a long-lived SIMP interacts with nucleon by the strength of the potential between two nucleons, the binding of the SIMP to a nucleon triggers formations of heavy nuclei up to carbon which are bound to the SIMP.  The decay lifetime of the SIMP, therefore, should be $\tau_X \lesssim 200$~s to avoid heavy element synthesis conflicting with observations of old stars \cite{Kusakabe:2009jt}.  

If the potential between a long-lived SIMP, i.e., $X$, and a nucleon is suitably weaker than that between two nucleons, primordial abundances of $^7$Be and $^7$Li can reduce by the reactions $X+^7$Be$\rightarrow ^3$He$+^4$He$_X$ and $X + ^7$Li$\rightarrow t+^4$He$_X$, respectively \cite{Kawasaki:2010yh}.  The discrepancy in $^7$Li abundance between predictions in standard big bang nucleosynthesis model and observations of metal-poor stars~\cite{Spite:1982dd,Ryan:2000zz,Melendez:2004ni,Asplund:2005yt,bon2007,Shi:2006zz,Aoki:2009ce} can be caused through the exotic nuclear reactions enabled by the existence of long-lived relic SIMPs.  The abundance of SIMPs necessary for a reduction of $^7$Li abundance down to the observed level in an example case (figures 5 and 7 of Ref.\cite{Kawasaki:2010yh}) was $\sim 1.7\times 10^{-4}$ times as large as the baryon abundance.  That relatively high SIMP abundance is predicted especially in Case 3 in this study.  The present study indicates that a high relic abundance tends to result from small branching ratios of $(Y\bar{Y})$ states for decays to lower $(Y\bar{Y})$ bound states, a large initial abundance of MCP, and large MCP mass.
Although we did not treat a case that the colored particle $Y$ is electrically neutral, its relic abundance may be larger than in charged cases as we have mentioned at the end of the Sec.~\ref{sec3}.  The neutral $Y$ may, therefore, also be a viable candidate for SIMP needed for a $^7$Li reduction.

There are interesting possible predictions to be tested in future astronomical observations adding to that on $^7$Li \cite{Kusakabe:2009jt}.  One is $^9$Be and/or B abundances larger than the prediction of standard big bang nucleosynthesis.  Another is a high isotopic ratio $^{10}$B/$^{11}$B different from predictions of other models for boron production such as the cosmic ray nucleosynthesis ($^{10}$B/$^{11}$B$\sim 0.4$~\cite{Prantzos1993,Ramaty1997,Kusakabe2008}) and the supernova neutrino process ($^{10}$B/$^{11}$B$\ll 1$~\cite{Woosley:1995ip,Yoshida:2005uy}).

If a long-lived SIMP affected the nucleosynthesis, primordial light
element abundances would have a unique information
on the quark-hadron phase transition in the early Universe.  The relic abundance of SIMP significantly depends on
hadronic characteristics through annihilation rates as shown in this
study.  In view of this, more elaborated studies on the annihilation
process are necessary to derive an information on the phase transition
in future.

\begin{acknowledgments}
This work is supported by Grant-in-Aid for Scientific Research from JSPS Grant No.21.6817 (Kusakabe).
\end{acknowledgments}

\bibliography{reference}

\begin{thebibliography}{47}
\expandafter\ifx\csname natexlab\endcsname\relax\def\natexlab#1{#1}\fi
\expandafter\ifx\csname bibnamefont\endcsname\relax
  \def\bibnamefont#1{#1}\fi
\expandafter\ifx\csname bibfnamefont\endcsname\relax
  \def\bibfnamefont#1{#1}\fi
\expandafter\ifx\csname citenamefont\endcsname\relax
  \def\citenamefont#1{#1}\fi
\expandafter\ifx\csname url\endcsname\relax
  \def\url#1{\texttt{#1}}\fi
\expandafter\ifx\csname urlprefix\endcsname\relax\def\urlprefix{URL }\fi
\providecommand{\bibinfo}[2]{#2}
\providecommand{\eprint}[2][]{\url{#2}}

\bibitem[{\citenamefont{Arkani-Hamed and
  Dimopoulos}(2005)}]{ArkaniHamed:2004fb}
\bibinfo{author}{\bibfnamefont{N.}~\bibnamefont{Arkani-Hamed}}
  \bibnamefont{and}
  \bibinfo{author}{\bibfnamefont{S.}~\bibnamefont{Dimopoulos}},
  \bibinfo{journal}{JHEP} \textbf{\bibinfo{volume}{06}}, \bibinfo{pages}{073}
  (\bibinfo{year}{2005}), \eprint{hep-th/0405159}.

\bibitem[{\citenamefont{Arkani-Hamed et~al.}(2005)\citenamefont{Arkani-Hamed,
  Dimopoulos, Giudice, and Romanino}}]{ArkaniHamed:2004yi}
\bibinfo{author}{\bibfnamefont{N.}~\bibnamefont{Arkani-Hamed}},
  \bibinfo{author}{\bibfnamefont{S.}~\bibnamefont{Dimopoulos}},
  \bibinfo{author}{\bibfnamefont{G.~F.} \bibnamefont{Giudice}},
  \bibnamefont{and} \bibinfo{author}{\bibfnamefont{A.}~\bibnamefont{Romanino}},
  \bibinfo{journal}{Nucl. Phys.} \textbf{\bibinfo{volume}{B709}},
  \bibinfo{pages}{3} (\bibinfo{year}{2005}), \eprint{hep-ph/0409232}.

\bibitem[{\citenamefont{Raby}(1998)}]{Raby:1997bpa}
\bibinfo{author}{\bibfnamefont{S.}~\bibnamefont{Raby}}, \bibinfo{journal}{Phys.
  Lett.} \textbf{\bibinfo{volume}{B422}}, \bibinfo{pages}{158}
  (\bibinfo{year}{1998}), \eprint{hep-ph/9712254}.

\bibitem[{\citenamefont{Shirai et~al.}(2010)\citenamefont{Shirai, Yamazaki, and
  Yonekura}}]{Shirai:2010rr}
\bibinfo{author}{\bibfnamefont{S.}~\bibnamefont{Shirai}},
  \bibinfo{author}{\bibfnamefont{M.}~\bibnamefont{Yamazaki}}, \bibnamefont{and}
  \bibinfo{author}{\bibfnamefont{K.}~\bibnamefont{Yonekura}},
  \bibinfo{journal}{JHEP} \textbf{\bibinfo{volume}{06}}, \bibinfo{pages}{056}
  (\bibinfo{year}{2010}), \eprint{1003.3155}.

\bibitem[{\citenamefont{Covi et~al.}(2010)\citenamefont{Covi, Olechowski,
  Pokorski, Turzynski, and Wells}}]{Covi:2010au}
\bibinfo{author}{\bibfnamefont{L.}~\bibnamefont{Covi}},
  \bibinfo{author}{\bibfnamefont{M.}~\bibnamefont{Olechowski}},
  \bibinfo{author}{\bibfnamefont{S.}~\bibnamefont{Pokorski}},
  \bibinfo{author}{\bibfnamefont{K.}~\bibnamefont{Turzynski}},
  \bibnamefont{and} \bibinfo{author}{\bibfnamefont{J.~D.} \bibnamefont{Wells}}
  (\bibinfo{year}{2010}), \eprint{1009.3801}.

\bibitem[{\citenamefont{Sarid and Thomas}(2000)}]{Sarid:1999zx}
\bibinfo{author}{\bibfnamefont{U.}~\bibnamefont{Sarid}} \bibnamefont{and}
  \bibinfo{author}{\bibfnamefont{S.~D.} \bibnamefont{Thomas}},
  \bibinfo{journal}{Phys. Rev. Lett.} \textbf{\bibinfo{volume}{85}},
  \bibinfo{pages}{1178} (\bibinfo{year}{2000}), \eprint{hep-ph/9909349}.

\bibitem[{\citenamefont{Hisano et~al.}(2010)\citenamefont{Hisano, Nakayama,
  Sugiyama, Takesako, and Yamanaka}}]{Hisano:2010bx}
\bibinfo{author}{\bibfnamefont{J.}~\bibnamefont{Hisano}},
  \bibinfo{author}{\bibfnamefont{K.}~\bibnamefont{Nakayama}},
  \bibinfo{author}{\bibfnamefont{S.}~\bibnamefont{Sugiyama}},
  \bibinfo{author}{\bibfnamefont{T.}~\bibnamefont{Takesako}}, \bibnamefont{and}
  \bibinfo{author}{\bibfnamefont{M.}~\bibnamefont{Yamanaka}},
  \bibinfo{journal}{Phys. Lett.} \textbf{\bibinfo{volume}{B691}},
  \bibinfo{pages}{46} (\bibinfo{year}{2010}), \eprint{1003.3648}.

\bibitem[{\citenamefont{Nakayama et~al.}(2010)\citenamefont{Nakayama,
  Takahashi, and Yanagida}}]{Nakayama:2010vs}
\bibinfo{author}{\bibfnamefont{K.}~\bibnamefont{Nakayama}},
  \bibinfo{author}{\bibfnamefont{F.}~\bibnamefont{Takahashi}},
  \bibnamefont{and} \bibinfo{author}{\bibfnamefont{T.~T.}
  \bibnamefont{Yanagida}} (\bibinfo{year}{2010}), \eprint{1010.5693}.

\bibitem[{\citenamefont{Wolfram}(1979)}]{Wolfram:1978gp}
\bibinfo{author}{\bibfnamefont{S.}~\bibnamefont{Wolfram}},
  \bibinfo{journal}{Phys. Lett.} \textbf{\bibinfo{volume}{B82}},
  \bibinfo{pages}{65} (\bibinfo{year}{1979}).

\bibitem[{\citenamefont{Dover et~al.}(1979)\citenamefont{Dover, Gaisser, and
  Steigman}}]{Dover:1979sn}
\bibinfo{author}{\bibfnamefont{C.~B.} \bibnamefont{Dover}},
  \bibinfo{author}{\bibfnamefont{T.~K.} \bibnamefont{Gaisser}},
  \bibnamefont{and} \bibinfo{author}{\bibfnamefont{G.}~\bibnamefont{Steigman}},
  \bibinfo{journal}{Phys. Rev. Lett.} \textbf{\bibinfo{volume}{42}},
  \bibinfo{pages}{1117} (\bibinfo{year}{1979}).

\bibitem[{\citenamefont{Starkman et~al.}(1990)\citenamefont{Starkman, Gould,
  Esmailzadeh, and Dimopoulos}}]{Starkman:1990nj}
\bibinfo{author}{\bibfnamefont{G.~D.} \bibnamefont{Starkman}},
  \bibinfo{author}{\bibfnamefont{A.}~\bibnamefont{Gould}},
  \bibinfo{author}{\bibfnamefont{R.}~\bibnamefont{Esmailzadeh}},
  \bibnamefont{and}
  \bibinfo{author}{\bibfnamefont{S.}~\bibnamefont{Dimopoulos}},
  \bibinfo{journal}{Phys. Rev.} \textbf{\bibinfo{volume}{D41}},
  \bibinfo{pages}{3594} (\bibinfo{year}{1990}).

\bibitem[{\citenamefont{{Nardi} and {Roulet}}(1990)}]{Nardi1990}
\bibinfo{author}{\bibfnamefont{E.}~\bibnamefont{{Nardi}}} \bibnamefont{and}
  \bibinfo{author}{\bibfnamefont{E.}~\bibnamefont{{Roulet}}},
  \bibinfo{journal}{Physics Letters B} \textbf{\bibinfo{volume}{245}},
  \bibinfo{pages}{105} (\bibinfo{year}{1990}).

\bibitem[{\citenamefont{Baer et~al.}(1999)\citenamefont{Baer, Cheung, and
  Gunion}}]{Baer:1998pg}
\bibinfo{author}{\bibfnamefont{H.}~\bibnamefont{Baer}},
  \bibinfo{author}{\bibfnamefont{K.-m.} \bibnamefont{Cheung}},
  \bibnamefont{and} \bibinfo{author}{\bibfnamefont{J.~F.}
  \bibnamefont{Gunion}}, \bibinfo{journal}{Phys. Rev.}
  \textbf{\bibinfo{volume}{D59}}, \bibinfo{pages}{075002}
  (\bibinfo{year}{1999}), \eprint{hep-ph/9806361}.

\bibitem[{\citenamefont{Kang et~al.}(2008)\citenamefont{Kang, Luty, and
  Nasri}}]{Kang:2006yd}
\bibinfo{author}{\bibfnamefont{J.}~\bibnamefont{Kang}},
  \bibinfo{author}{\bibfnamefont{M.~A.} \bibnamefont{Luty}}, \bibnamefont{and}
  \bibinfo{author}{\bibfnamefont{S.}~\bibnamefont{Nasri}},
  \bibinfo{journal}{JHEP} \textbf{\bibinfo{volume}{09}}, \bibinfo{pages}{086}
  (\bibinfo{year}{2008}), \eprint{hep-ph/0611322}.

\bibitem[{\citenamefont{{Kolb} and {Turner}}(1990)}]{kolb1990}
\bibinfo{author}{\bibfnamefont{E.~W.} \bibnamefont{{Kolb}}} \bibnamefont{and}
  \bibinfo{author}{\bibfnamefont{M.~S.} \bibnamefont{{Turner}}},
  \emph{\bibinfo{title}{{The early universe}}}
  (\bibinfo{publisher}{Addison-Wesley, Reading, MA}, \bibinfo{year}{1990}).

\bibitem[{\citenamefont{Diaz-Cruz et~al.}(2007)\citenamefont{Diaz-Cruz, Ellis,
  Olive, and Santoso}}]{DiazCruz:2007fc}
\bibinfo{author}{\bibfnamefont{J.~L.} \bibnamefont{Diaz-Cruz}},
  \bibinfo{author}{\bibfnamefont{J.~R.} \bibnamefont{Ellis}},
  \bibinfo{author}{\bibfnamefont{K.~A.} \bibnamefont{Olive}}, \bibnamefont{and}
  \bibinfo{author}{\bibfnamefont{Y.}~\bibnamefont{Santoso}},
  \bibinfo{journal}{JHEP} \textbf{\bibinfo{volume}{05}}, \bibinfo{pages}{003}
  (\bibinfo{year}{2007}), \eprint{hep-ph/0701229}.

\bibitem[{\citenamefont{Jacoby and Nussinov}(2007)}]{Jacoby:2007nw}
\bibinfo{author}{\bibfnamefont{C.}~\bibnamefont{Jacoby}} \bibnamefont{and}
  \bibinfo{author}{\bibfnamefont{S.}~\bibnamefont{Nussinov}}
  (\bibinfo{year}{2007}), \eprint{0712.2681}.

\bibitem[{\citenamefont{Nussinov and Jacoby}(2009)}]{Nussinov:2009hc}
\bibinfo{author}{\bibfnamefont{S.}~\bibnamefont{Nussinov}} \bibnamefont{and}
  \bibinfo{author}{\bibfnamefont{C.}~\bibnamefont{Jacoby}}
  (\bibinfo{year}{2009}), \eprint{0907.4932}.

\bibitem[{\citenamefont{Okun}(1980)}]{Okun:1980kw}
\bibinfo{author}{\bibfnamefont{L.~B.} \bibnamefont{Okun}},
  \bibinfo{journal}{JETP Lett.} \textbf{\bibinfo{volume}{31}},
  \bibinfo{pages}{144} (\bibinfo{year}{1980}).

\bibitem[{\citenamefont{{Okun}}(1980)}]{Okun1980b}
\bibinfo{author}{\bibfnamefont{L.~B.} \bibnamefont{{Okun}}},
  \bibinfo{journal}{Nuclear Physics B} \textbf{\bibinfo{volume}{173}},
  \bibinfo{pages}{1} (\bibinfo{year}{1980}).

\bibitem[{\citenamefont{Gupta and Quinn}(1982)}]{Gupta:1981ve}
\bibinfo{author}{\bibfnamefont{S.}~\bibnamefont{Gupta}} \bibnamefont{and}
  \bibinfo{author}{\bibfnamefont{H.~R.} \bibnamefont{Quinn}},
  \bibinfo{journal}{Phys. Rev.} \textbf{\bibinfo{volume}{D25}},
  \bibinfo{pages}{838} (\bibinfo{year}{1982}).

\bibitem[{\citenamefont{Kang and Luty}(2009)}]{Kang:2008ea}
\bibinfo{author}{\bibfnamefont{J.}~\bibnamefont{Kang}} \bibnamefont{and}
  \bibinfo{author}{\bibfnamefont{M.~A.} \bibnamefont{Luty}},
  \bibinfo{journal}{JHEP} \textbf{\bibinfo{volume}{11}}, \bibinfo{pages}{065}
  (\bibinfo{year}{2009}), \eprint{0805.4642}.

\bibitem[{\citenamefont{Strassler and Zurek}(2007)}]{Strassler:2006im}
\bibinfo{author}{\bibfnamefont{M.~J.} \bibnamefont{Strassler}}
  \bibnamefont{and} \bibinfo{author}{\bibfnamefont{K.~M.} \bibnamefont{Zurek}},
  \bibinfo{journal}{Phys. Lett.} \textbf{\bibinfo{volume}{B651}},
  \bibinfo{pages}{374} (\bibinfo{year}{2007}), \eprint{hep-ph/0604261}.

\bibitem[{\citenamefont{{Richard-Serre}
  et~al.}(1970)\citenamefont{{Richard-Serre}, {Hirt}, {Measday}, {Michaelis},
  {Saltmarsh}, and {Skarek}}}]{Richard1970}
\bibinfo{author}{\bibfnamefont{C.}~\bibnamefont{{Richard-Serre}}},
  \bibinfo{author}{\bibfnamefont{W.}~\bibnamefont{{Hirt}}},
  \bibinfo{author}{\bibfnamefont{D.~F.} \bibnamefont{{Measday}}},
  \bibinfo{author}{\bibfnamefont{E.~G.} \bibnamefont{{Michaelis}}},
  \bibinfo{author}{\bibfnamefont{M.~J.~M.} \bibnamefont{{Saltmarsh}}},
  \bibnamefont{and} \bibinfo{author}{\bibfnamefont{P.}~\bibnamefont{{Skarek}}},
  \bibinfo{journal}{Nuclear Physics B} \textbf{\bibinfo{volume}{20}},
  \bibinfo{pages}{413} (\bibinfo{year}{1970}).

\bibitem[{\citenamefont{{Mandelstam}}(1958)}]{Mandelstam1958}
\bibinfo{author}{\bibfnamefont{S.}~\bibnamefont{{Mandelstam}}},
  \bibinfo{journal}{Royal Society of London Proceedings Series A}
  \textbf{\bibinfo{volume}{244}}, \bibinfo{pages}{491} (\bibinfo{year}{1958}).

\bibitem[{\citenamefont{{Schiff} and {Tran Thanh van}}(1968)}]{Schiff1968}
\bibinfo{author}{\bibfnamefont{D.}~\bibnamefont{{Schiff}}} \bibnamefont{and}
  \bibinfo{author}{\bibfnamefont{J.}~\bibnamefont{{Tran Thanh van}}},
  \bibinfo{journal}{Nuclear Physics B} \textbf{\bibinfo{volume}{5}},
  \bibinfo{pages}{529} (\bibinfo{year}{1968}).

\bibitem[{\citenamefont{{Clayton}}(1983)}]{clayton-book}
\bibinfo{author}{\bibfnamefont{D.~D.} \bibnamefont{{Clayton}}},
  \emph{\bibinfo{title}{{Principles of stellar evolution and nucleosynthesis}}}
  (\bibinfo{publisher}{University of Chicago Press, Chicago, IL},
  \bibinfo{year}{1983}).

\bibitem[{\citenamefont{{Nakamura} and {Particle Data Group}}(2010)}]{PDG2010}
\bibinfo{author}{\bibfnamefont{K.}~\bibnamefont{{Nakamura}}} \bibnamefont{and}
  \bibinfo{author}{\bibnamefont{{Particle Data Group}}},
  \bibinfo{journal}{Journal of Physics G Nuclear Physics}
  \textbf{\bibinfo{volume}{37}}, \bibinfo{pages}{075021}
  (\bibinfo{year}{2010}).

\bibitem[{\citenamefont{{Pagel}}(1997)}]{pagel1997book}
\bibinfo{author}{\bibfnamefont{B.~E.~J.} \bibnamefont{{Pagel}}},
  \emph{\bibinfo{title}{{Nucleosynthesis and Chemical Evolution of Galaxies}}}
  (\bibinfo{publisher}{Cambridge University Press, Cambridge, England},
  \bibinfo{year}{1997}).

\bibitem[{\citenamefont{Brambilla et~al.}(2011)}]{Brambilla:2010cs}
\bibinfo{author}{\bibfnamefont{N.}~\bibnamefont{Brambilla}}
  \bibnamefont{et~al.}, \bibinfo{journal}{Eur. Phys. J.}
  \textbf{\bibinfo{volume}{C71}}, \bibinfo{pages}{1534} (\bibinfo{year}{2011}),
  \eprint{1010.5827}.

\bibitem[{\citenamefont{Mackeprang and Milstead}(2010)}]{Mackeprang:2009ad}
\bibinfo{author}{\bibfnamefont{R.}~\bibnamefont{Mackeprang}} \bibnamefont{and}
  \bibinfo{author}{\bibfnamefont{D.}~\bibnamefont{Milstead}},
  \bibinfo{journal}{Eur. Phys. J.} \textbf{\bibinfo{volume}{C66}},
  \bibinfo{pages}{493} (\bibinfo{year}{2010}), \eprint{0908.1868}.

\bibitem[{\citenamefont{{Gondolo} and {Gelmini}}(1991)}]{Gondolo1991}
\bibinfo{author}{\bibfnamefont{P.}~\bibnamefont{{Gondolo}}} \bibnamefont{and}
  \bibinfo{author}{\bibfnamefont{G.}~\bibnamefont{{Gelmini}}},
  \bibinfo{journal}{Nuclear Physics B} \textbf{\bibinfo{volume}{360}},
  \bibinfo{pages}{145} (\bibinfo{year}{1991}).

\bibitem[{\citenamefont{Kusakabe et~al.}(2009)\citenamefont{Kusakabe, Kajino,
  Yoshida, and Mathews}}]{Kusakabe:2009jt}
\bibinfo{author}{\bibfnamefont{M.}~\bibnamefont{Kusakabe}},
  \bibinfo{author}{\bibfnamefont{T.}~\bibnamefont{Kajino}},
  \bibinfo{author}{\bibfnamefont{T.}~\bibnamefont{Yoshida}}, \bibnamefont{and}
  \bibinfo{author}{\bibfnamefont{G.~J.} \bibnamefont{Mathews}},
  \bibinfo{journal}{Phys. Rev.} \textbf{\bibinfo{volume}{D80}},
  \bibinfo{pages}{103501} (\bibinfo{year}{2009}), \eprint{0906.3516}.

\bibitem[{\citenamefont{Kawasaki and Kusakabe}(2011)}]{Kawasaki:2010yh}
\bibinfo{author}{\bibfnamefont{M.}~\bibnamefont{Kawasaki}} \bibnamefont{and}
  \bibinfo{author}{\bibfnamefont{M.}~\bibnamefont{Kusakabe}},
  \bibinfo{journal}{Phys. Rev.} \textbf{\bibinfo{volume}{D83}},
  \bibinfo{pages}{055011} (\bibinfo{year}{2011}), \eprint{1012.0435}.

\bibitem[{\citenamefont{Spite and Spite}(1982)}]{Spite:1982dd}
\bibinfo{author}{\bibfnamefont{F.}~\bibnamefont{Spite}} \bibnamefont{and}
  \bibinfo{author}{\bibfnamefont{M.}~\bibnamefont{Spite}},
  \bibinfo{journal}{Astron. Astrophys.} \textbf{\bibinfo{volume}{115}},
  \bibinfo{pages}{357} (\bibinfo{year}{1982}).

\bibitem[{\citenamefont{Ryan et~al.}(2000)\citenamefont{Ryan, Beers, Olive,
  Fields, and Norris}}]{Ryan:2000zz}
\bibinfo{author}{\bibfnamefont{S.~G.} \bibnamefont{Ryan}},
  \bibinfo{author}{\bibfnamefont{T.~C.} \bibnamefont{Beers}},
  \bibinfo{author}{\bibfnamefont{K.~A.} \bibnamefont{Olive}},
  \bibinfo{author}{\bibfnamefont{B.~D.} \bibnamefont{Fields}},
  \bibnamefont{and} \bibinfo{author}{\bibfnamefont{J.~E.}
  \bibnamefont{Norris}}, \bibinfo{journal}{Astrophys. J.}
  \textbf{\bibinfo{volume}{530}}, \bibinfo{pages}{L57} (\bibinfo{year}{2000}).

\bibitem[{\citenamefont{Melendez and Ramirez}(2004)}]{Melendez:2004ni}
\bibinfo{author}{\bibfnamefont{J.}~\bibnamefont{Melendez}} \bibnamefont{and}
  \bibinfo{author}{\bibfnamefont{I.}~\bibnamefont{Ramirez}},
  \bibinfo{journal}{Astrophys. J.} \textbf{\bibinfo{volume}{615}},
  \bibinfo{pages}{L33} (\bibinfo{year}{2004}), \eprint{astro-ph/0409383}.

\bibitem[{\citenamefont{Asplund et~al.}(2006)\citenamefont{Asplund, Lambert,
  Nissen, Primas, and Smith}}]{Asplund:2005yt}
\bibinfo{author}{\bibfnamefont{M.}~\bibnamefont{Asplund}},
  \bibinfo{author}{\bibfnamefont{D.~L.} \bibnamefont{Lambert}},
  \bibinfo{author}{\bibfnamefont{P.~E.} \bibnamefont{Nissen}},
  \bibinfo{author}{\bibfnamefont{F.}~\bibnamefont{Primas}}, \bibnamefont{and}
  \bibinfo{author}{\bibfnamefont{V.~V.} \bibnamefont{Smith}},
  \bibinfo{journal}{Astrophys. J.} \textbf{\bibinfo{volume}{644}},
  \bibinfo{pages}{229} (\bibinfo{year}{2006}), \eprint{astro-ph/0510636}.

\bibitem[{\citenamefont{{Bonifacio} et~al.}(2007)\citenamefont{{Bonifacio},
  {Molaro}, {Sivarani}, {Cayrel}, {Spite}, {Spite}, {Plez}, {Andersen},
  {Barbuy}, {Beers} et~al.}}]{bon2007}
\bibinfo{author}{\bibfnamefont{P.}~\bibnamefont{{Bonifacio}}},
  \bibinfo{author}{\bibfnamefont{P.}~\bibnamefont{{Molaro}}},
  \bibinfo{author}{\bibfnamefont{T.}~\bibnamefont{{Sivarani}}},
  \bibinfo{author}{\bibfnamefont{R.}~\bibnamefont{{Cayrel}}},
  \bibinfo{author}{\bibfnamefont{M.}~\bibnamefont{{Spite}}},
  \bibinfo{author}{\bibfnamefont{F.}~\bibnamefont{{Spite}}},
  \bibinfo{author}{\bibfnamefont{B.}~\bibnamefont{{Plez}}},
  \bibinfo{author}{\bibfnamefont{J.}~\bibnamefont{{Andersen}}},
  \bibinfo{author}{\bibfnamefont{B.}~\bibnamefont{{Barbuy}}},
  \bibinfo{author}{\bibfnamefont{T.~C.} \bibnamefont{{Beers}}},
  \bibnamefont{et~al.}, \bibinfo{journal}{Astron. Astrophys.}
  \textbf{\bibinfo{volume}{462}}, \bibinfo{pages}{851} (\bibinfo{year}{2007}),
  \eprint{arXiv:astro-ph/0610245}.

\bibitem[{\citenamefont{Shi et~al.}(2007)\citenamefont{Shi, Gehren, Zhang,
  Zeng, and Zhao}}]{Shi:2006zz}
\bibinfo{author}{\bibfnamefont{J.~R.} \bibnamefont{Shi}},
  \bibinfo{author}{\bibfnamefont{T.}~\bibnamefont{Gehren}},
  \bibinfo{author}{\bibfnamefont{H.~W.} \bibnamefont{Zhang}},
  \bibinfo{author}{\bibfnamefont{J.~L.} \bibnamefont{Zeng}}, \bibnamefont{and}
  \bibinfo{author}{\bibfnamefont{G.}~\bibnamefont{Zhao}},
  \bibinfo{journal}{Astron. Astrophys.} \textbf{\bibinfo{volume}{465}},
  \bibinfo{pages}{587} (\bibinfo{year}{2007}).

\bibitem[{\citenamefont{Aoki et~al.}(2009)}]{Aoki:2009ce}
\bibinfo{author}{\bibfnamefont{W.}~\bibnamefont{Aoki}} \bibnamefont{et~al.},
  \bibinfo{journal}{Astrophys. J.} \textbf{\bibinfo{volume}{698}},
  \bibinfo{pages}{1803} (\bibinfo{year}{2009}), \eprint{0904.1448}.

\bibitem[{\citenamefont{{Prantzos} et~al.}(1993)\citenamefont{{Prantzos},
  {Casse}, and {Vangioni-Flam}}}]{Prantzos1993}
\bibinfo{author}{\bibfnamefont{N.}~\bibnamefont{{Prantzos}}},
  \bibinfo{author}{\bibfnamefont{M.}~\bibnamefont{{Casse}}}, \bibnamefont{and}
  \bibinfo{author}{\bibfnamefont{E.}~\bibnamefont{{Vangioni-Flam}}},
  \bibinfo{journal}{\apj} \textbf{\bibinfo{volume}{403}}, \bibinfo{pages}{630}
  (\bibinfo{year}{1993}).

\bibitem[{\citenamefont{{Ramaty} et~al.}(1997)\citenamefont{{Ramaty},
  {Kozlovsky}, {Lingenfelter}, and {Reeves}}}]{Ramaty1997}
\bibinfo{author}{\bibfnamefont{R.}~\bibnamefont{{Ramaty}}},
  \bibinfo{author}{\bibfnamefont{B.}~\bibnamefont{{Kozlovsky}}},
  \bibinfo{author}{\bibfnamefont{R.~E.} \bibnamefont{{Lingenfelter}}},
  \bibnamefont{and} \bibinfo{author}{\bibfnamefont{H.}~\bibnamefont{{Reeves}}},
  \bibinfo{journal}{Astrophys. J.} \textbf{\bibinfo{volume}{488}},
  \bibinfo{pages}{730} (\bibinfo{year}{1997}), \eprint{arXiv:astro-ph/9610255}.

\bibitem[{\citenamefont{{Kusakabe}}(2008)}]{Kusakabe2008}
\bibinfo{author}{\bibfnamefont{M.}~\bibnamefont{{Kusakabe}}},
  \bibinfo{journal}{Astrophys. J.} \textbf{\bibinfo{volume}{681}},
  \bibinfo{pages}{18} (\bibinfo{year}{2008}), \eprint{0803.3401}.

\bibitem[{\citenamefont{Woosley and Weaver}(1995)}]{Woosley:1995ip}
\bibinfo{author}{\bibfnamefont{S.~E.} \bibnamefont{Woosley}} \bibnamefont{and}
  \bibinfo{author}{\bibfnamefont{T.~A.} \bibnamefont{Weaver}},
  \bibinfo{journal}{Astrophys. J. Suppl.} \textbf{\bibinfo{volume}{101}},
  \bibinfo{pages}{181} (\bibinfo{year}{1995}).

\bibitem[{\citenamefont{Yoshida et~al.}(2005)\citenamefont{Yoshida, Kajino, and
  Hartmann}}]{Yoshida:2005uy}
\bibinfo{author}{\bibfnamefont{T.}~\bibnamefont{Yoshida}},
  \bibinfo{author}{\bibfnamefont{T.}~\bibnamefont{Kajino}}, \bibnamefont{and}
  \bibinfo{author}{\bibfnamefont{D.~H.} \bibnamefont{Hartmann}},
  \bibinfo{journal}{Phys. Rev. Lett.} \textbf{\bibinfo{volume}{94}},
  \bibinfo{pages}{231101} (\bibinfo{year}{2005}), \eprint{astro-ph/0505043}.

\bibitem[{\citenamefont{Wong}(2002)}]{Wong:2001uu}
\bibinfo{author}{\bibfnamefont{C.-Y.} \bibnamefont{Wong}},
  \bibinfo{journal}{Phys. Rev.} \textbf{\bibinfo{volume}{C65}},
  \bibinfo{pages}{034902} (\bibinfo{year}{2002}), \eprint{nucl-th/0110004}.

\end{thebibliography}

%%%%%%%%%%%%%%%%%%%%%%%%%%%%%%%%%%%%%%%%%%%%%%%%%%%%%%
%\begin{thebibliography}{99}
%%%%%%%%%%%%%%%%%%%%%%%%%%%%%%%%%%%%%%%%%%%%%%%%%%%%%%

%\end{thebibliography}

\end{document}